\documentclass[12pt, draftclsnofoot, onecolumn]{IEEEtran}
\usepackage{stmaryrd}
\usepackage{amsmath}
\usepackage{amssymb}
\usepackage{psfrag}
\usepackage{graphicx}
\usepackage{epstopdf}
\usepackage{todonotes}
\usepackage{multirow}
\usepackage{booktabs}
\usepackage{siunitx}
\usepackage{cite}
\usepackage{algorithmic}
\usepackage{algorithm}
\usepackage{svg}
\usepackage{mathtools}
\usepackage{hyperref}

\usepackage{float}
\ifCLASSOPTIONcompsoc
\usepackage[caption=false,font=normalsize,labelfont=sf,textfont=sf]{subfig}
\else
\usepackage[caption=false,font=footnotesize]{subfig}
\fi

\title{Learning to Equalize OTFS}
\begin{document}
\author{\IEEEauthorblockN{Zhou Zhou, Lingjia Liu, Jiarui Xu, and Robert Calderbank}
\thanks{Z. Zhou, L. Liu, and J. Xu are with Wireless@VT, Bradley Department of Electrical and Computer Engineering, Virginia Tech, USA. R. Calderbank is with Department of Electrical and Computer Engineering, Duke University, USA. 
The work of Z. Zhou, L. Liu, J. Xu and R. Calderbank is supported in part by the AFRL/AFOSR University Center of Excellence (COE) under the grant FA 8750-20-2-0504. The work of R. Calderbank is also supported in part by AFRL under the grant FA 9550-20-1-0266. 
}}
\maketitle
\begin{abstract}
Orthogonal Time Frequency Space (OTFS) is a novel framework that processes modulation symbols via a time-independent channel characterized by the delay-Doppler domain. The conventional waveform, orthogonal frequency division multiplexing (OFDM), requires tracking frequency selective fading channels over the time, whereas OTFS benefits from full time-frequency diversity by leveraging appropriate equalization techniques. 
In this paper, we consider a neural network-based supervised learning framework for OTFS equalization. 
Learning of the introduced neural network is conducted in each OTFS frame fulfilling an online learning framework: the training and testing datasets are within the same OTFS-frame over the air. 
Utilizing reservoir computing, a special recurrent neural network, the resulting one-shot online learning is sufficiently flexible to cope with channel variations among different OTFS frames (e.g., due to the link/rank adaptation and user scheduling in cellular networks). The proposed method does not require explicit channel state information (CSI) and simulation results demonstrate a lower bit error rate (BER) than conventional equalization methods in the low signal-to-noise (SNR) regime under large Doppler spreads. When compared with its neural network-based counterparts for OFDM, the introduced approach for OTFS will lead to a better tradeoff between the processing complexity and the equalization performance. 
\end{abstract}
\section{Introduction}
Telecommunication vendors are expanding their services to various new scenarios, such as the Internet of Things (IoT),  the vehicle-to-everything (V2X), and the global coverage by non-terrestrial networks (NTN) including both high altitude platforms (HAPs) and low earth orbit (LEO) satellites in the 5G-Advanced era \cite{6G_white_paper}. 
For many of those high mobility scenarios such as V2X and NTN, it is imperative to ask: Will the current 5G air-interface design be sufficient for these new scenarios? Will 5G-advanced and 6G benefit from applying an alternative waveform design to accommodate these new scenarios? 
The lens provided by Orthogonal Time Frequency Space (OTFS) modulation~\cite{hadani2017orthogonal} has shed light on these questions. 
This approach circumvents the fundamental bottlenecks in the time-frequency processing framework of Orthogonal Frequency Division Modulation (OFDM) which is the key physical layer waveform of 4G LTE-Advanced~\cite{4GMIMO_OFDM} and 5G NR~\cite{5GAI}. Meanwhile, OTFS is a signal design framework/control plane that improves network layer functions like scheduling that are essential to realizing the promise of 5G/6G.  
As an alternative to OFDM, OTFS is rooted in an ``invariant expression'' of the doubly-selective wireless channel via the delay-Doppler domain~\cite{hadani2017orthogonal}. The resulting channel impulse response exhibits sparsity and persists for longer time than projections in time or frequency. These channel features enable the algorithm and protocol design of OTFS systems in channel estimation/tracking and equalization with reduced signaling overhead. 
In addition to the equalization benefits, it also can offer waveform benefits: the cyclic prefix (CP) of each OFDM symbol can be completely abolished in OTFS, significantly improving spectral efficiency especially for high mobility scenarios where CP is a significant system overhead for OFDM systems.

For high mobility scenarios with high Doppler shifts, transmitted symbols residing in the delay-Doppler domain experience a channel that changes more slowly compared with counterparts in the time-frequency domain~\cite{hadani2018otfs}. 
OTFS is implemented by allocating modulation symbols in the delay-Doppler domain. 
Accordingly, the time-varying multi-path wireless channel becomes a near-constant impulse response in the delay-Doppler domain via the symplectic Fourier transformation (i.e., Zak transform). 
Channel representation in the delay-Doppler domain expresses slow variation since channel reflective scatters are relatively static in the surrounding environment. On the other hand, OTFS can be viewed as a spreading modulation scheme in the time-frequency domain, where the basis functions are selected from a two-dimensional Fourier series. In this perspective, OTFS can be regarded as a 2D-CDMA scheme, whereas OFDM is essentially a one-dimensional spreading scheme using a single-mode Fourier basis. 
Furthermore, OTFS can achieve enhanced channel diversity associated with the use of appropriate equalization methods~\cite{hadani2018otfs, low_complex}.

\subsection{Signal Processing Challenges in OTFS Systems}
\label{sec_challenges}
In OTFS, the channel is modeled as a 2D multi-path response spread over the dimension of the delay-tap and the Doppler-shift. 
Hence, equalization techniques at the receiver can be introduced to exploit the diversity over these two dimensions. 
However, the equalizer has to address the following practical challenges:
\begin{itemize}
\item {\textbf{Challenge 1: Pilot Overhead}.} Conducting equalization in the Delay-Doppler domain requires pilots or reference signals for either explicit channel state information (CSI) estimation or equalizer coefficients adaptation. 
Due to the multi-tap nature of the delay-Doppler channel representation, guard intervals between pilot and data symbols are often inserted to avoid the underlying interference~\cite{raviteja2019embedded}. 
When OTFS systems are configured with multiple-input-and-multiple-output (MIMO) antennas, the overhead resulting from the guard intervals becomes a bottleneck, especially for massive MIMO systems. 
Therefore, it is important to resolve the pilot overhead issue by designing pilot patterns that support associated low complexity equalization methods.
\item {\textbf{Challenge 2: Accurate Channel Knowledge}.} Acquiring accurate CSI in the delay-Doppler domain is critical for model-based equalizers in OTFS systems. Although time-varying multi-path channels can be ideally approximated as a constant impulse response in the delay-Doppler domain, high received SNRs of the corresponding pilots or reference signals are needed to acquire accurate 2D channel spreading profiles. 
However, obtaining high received SNRs for pilots or reference signals is challenging in OTFS systems due to power and hardware constraints.
For example, when synchronization is not perfect, the transmitting and receiving pulse shaping filters in OTFS will fail to be biorthogonal. This introduces additional inter-carrier interference (ICI) \footnote{ICI happens when the OTFS system is implemented as an overlay to OFDM systems which will be discussed in Sec. \ref{c_ofdm}} and inter-symbol interference (ISI)~\cite{biglieri2019error} leading to low received SNRs.
Furthermore, the motion of channel scatters can cause misalignment to the channel-prior assuming a strict sparsity in the delay-Doppler domain. 
Therefore, it will result in a model mismatch to sparsity-based channel estimation algorithms.
\end{itemize}

\subsection{Related Work}
The equalization of OFDM in doubly selective channels has been extensively studied. 
In \cite{rugini2005simple}, a low complexity minimum mean-squared error (MMSE) based equalization method for OFDM systems over a time-varying channel has been introduced. 
A two-stage MMSE equalization method was considered in~\cite{schniter2004low} where the first step is to restrict the support of inter-carrier interference followed by eliminating the interference using successive MMSE. 
However, all these signal processing techniques rely on explicit accurate CSI. 
It is important to note that OFDM can be viewed as the projection of OTFS onto the time-frequency domain.
In fact, this projection will change more abruptly making the tracking of channel more challenging. 
For example, the above introduced OFDM-based schemes require channel tracking operating on a sub-millisecond basis under high mobility environments for 5G NR networks.
This makes it difficult to adopt neural network-based receive processing strategies. 

To handle wireless channels with high Doppler shifts, state-of-the-art equalization methods in OTFS systems often assume explicit CSI in the delay-Doppler domain is ideally available~\cite{hadani2017orthogonal, hadani2018otfs}. 
On the other hand, various channel estimation methods have been introduced to estimate explicit CSI using the pilot/reference signals.
For example, the work in \cite{channel_est} introduced a channel estimation method with almost linear complexity. 
Embedded pilot-aided channel estimation schemes for OTFS are investigated in~\cite{raviteja2019embedded}. 
In each OTFS frame, guard intervals are placed between pilot and data symbols in the delay-Doppler domain. 
The underlying CSI is estimated by performing hard thresholding and the estimated CSI is utilized for symbol detection. 
Note that the scenario considered is relatively simple as each delay tap is associated with single Doppler shift generated using Jakes’ formula. 
An alternative to using impulse signals as channel estimation pilots in the delay-Doppler domain is to use pseudo-noise (PN) sequences in the delay-Doppler domain~\cite{murali2018otfs}. 

Assuming ideal explicit CSI is available, a message-passing-based iterative algorithm for equalization in the delay-Doppler domain is introduced in \cite{raviteja2018interference}.
Note that the extension to MIMO-OTFS systems has not yet been explored.
Reference~\cite{thaj2020low} introduces a rake receiver in conjunction with decision feedback to conduct the symbol detection.
Ideal channel knowledge is assumed for turbo-boosting the success rate. 
Similar concepts of leveraging decision feedback have been widely studied in MIMO systems with multi-path channels, see~\cite{komninakis2002multi, al2000finite, duel1992equalizers}. However, feedback errors can propagate to amplify the detection error especially in the low SNR regime. When these methods are generalized to equalization of OTFS in the 2D delay-Doppler domain, their efficiency has yet to be studied. 

\subsection{One-Shot Online Learning for OTFS through Reservoir Computing}
To address the above mentioned challenges for OTFS systems, we consider using neural networks (NNs) to conduct equalization. 
In 4G and 5G networks where OFDM is used as the underlying waveform, there exist many MIMO transmission modes with link adaptation, rank adaptation, and scheduling operating on a subframe basis~\cite{4GMIMO_OFDM}. 
For high mobility scenarios that are of interest to 5G-Advanced and 6G, it is critical to design an online learning-based equalization method for OTFS systems that is adaptive and robust to the change of wireless environments, such as channel distributions, operation modes, scheduling decisions, and inter-user interference.
However, wireless networks, especially cellular networks, have unique features and constraints. 
For example, the online over-the-air (OTA) training dataset (e.g., reference signals) of a cellular network is extremely limited since it is the system rather than the data.
In 5G NR, the overhead defined for the demodulation reference signal is at most  20\%~\cite{5GNR1}. 
Therefore, NN-based online equalization for OTFS with very limited OTA training datasets is extremely challenging. 
To address this challenge, we consider a ``one-shot learning'' solution: In contrast to most learning-based algorithms, which require training on extensive datasets, one-shot learning aims to learn information from very few training samples. 
Intuitively, the superior generalization performance of the one-shot learning concept is achieved by incorporating NNs with inductive prior of unseen data \cite{miller2000learning}.

This paper considers using reservoir computing (RC, a special recurrent neural network where the structure is embedded with a sequential inductive bias) for equalization in OTFS systems. 
RC has been previously utilized for spatial and temporal interference cancellation tasks in MIMO-OFDM systems \cite{Mosleh2018RC,zhou2020learning,zhou2020deep,zhou2021RCnet}, and has proven to be efficient in handling high Doppler shift in an online fashion even with extremely limited OTA training datasets.
In the RC framework for MIMO-OFDM, the temporal-spatial signal features are read out to conduct interference cancellation in consecutive OFDM symbols.
In this paper, we generalize the concept by training the RC using a subset of received symbols (designed as specific pilot patterns) and applying the trained network on the remaining symbols (to equalize interference residing in data symbols). 
Our technical contributions in applying RC to equalization in OTFS are the following
\begin{itemize}
\item {\textbf{Novel Pilot Structures}:} We investigate two OTFS pilot structures in the delay-Doppler domain: interleaved pilot and superimposed pilot. Different from conventional pilot designs, the considered pilot and data symbols are allowed to interfere with each other without guard intervals between them. 
The training framework only utilizes pilot symbols within the same OTFS frame as the data symbols instead of those in previous OTFS frames or offline training datasets. 
Therefore, it offers a one-shot learning objective in terms of optimizing the NN weights learning through each individual OTFS frame. 
Since our design and approach is guard interval free and a purely online-learning based approach, it can effectively address \emph{Challenge 1} listed in Sec. \ref{sec_challenges}. 
Furthermore, the pilot overhead in our approach can be adaptively adjusted, providing a set of feasible configurations in the trade-off space of control overhead and system performance.
    
\item {\textbf{Robust Learning Algorithms}:} We develop novel RC-based learning algorithms for the two types of pilot pattern. 
In the first pilot structure, the learning algorithms can jointly cope with the interference between the pilot and data symbols regarding the interleaved pilot pattern. 
By contrast, it is the signal power split between pilot and data for the superimposed pilot that enables the equalization NN to generalize. More important, the learning algorithm does not rely on explicit CSI, providing robust equalization performance, even in low SNR regimes. 
This provides us an effective method to address \emph{Challenge 2} raised in Sec. \ref{sec_challenges}. 
\end{itemize}

The remainder of this paper is organized as follows: In Section \ref{prelim}, we briefly introduce the preliminaries of OTFS systems. In Section \ref{learning_method}, we develop a learning-based framework for OTFS equalization in the delay-Doppler domain. 
Our discussion of systems considerations includes the pilot design (training dataset), the neural network structure, and the loss objective for the equalization. In Section \ref{Ext}, we extend the learning framework to MIMO-OTFS. We also introduce an approach where we apply multiple RCs in the equalization training of each OTFS frame. Section \ref{evaluation} presents the comparison between conventional equalization methods and our proposed  method in both SISO and MIMO scenarios. We conclude the paper in Section \ref{conclusion}.

\section{Preliminaries}
\label{prelim}
A wireless channel exhibiting both time and frequency selectivity is referred to as a doubly-selective channel. This section first provides a mathematical representation of wireless communications signals over the doubly-selective channel. Then, we discuss the OTFS modulation and its connection to the OFDM system. Finally, we present the end-to-end relation between transmitted symbols and received symbols in the OTFS system in the delay-Doppler domain. 

\subsection{Doubly-Selective Channel}
\begin{figure}[!h]
    \centering
	\subfloat[]{\includegraphics[width=0.65\linewidth]{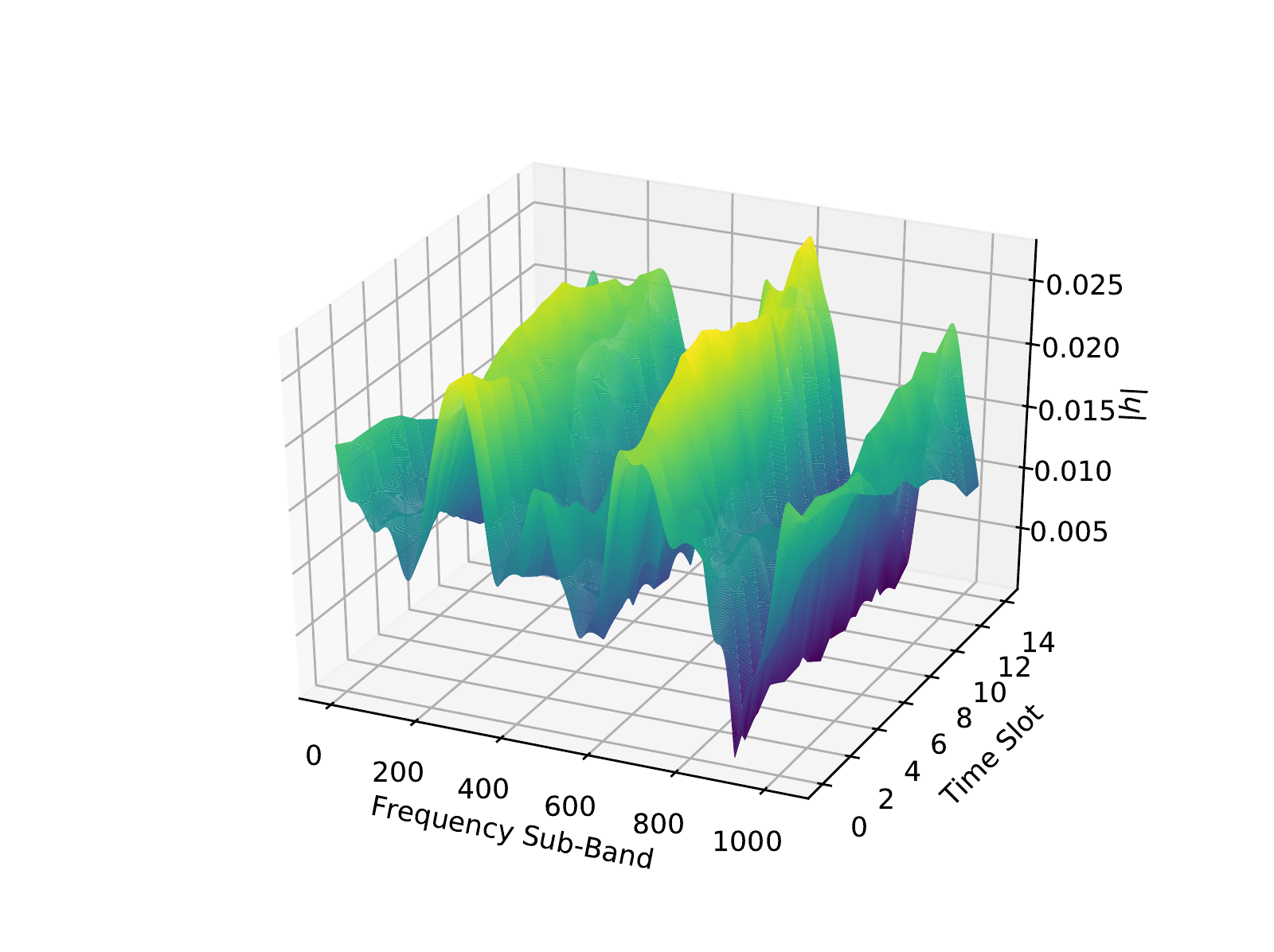}
		\label{tf}}
	\vfil
	\subfloat[]{\includegraphics[width=0.65\linewidth]{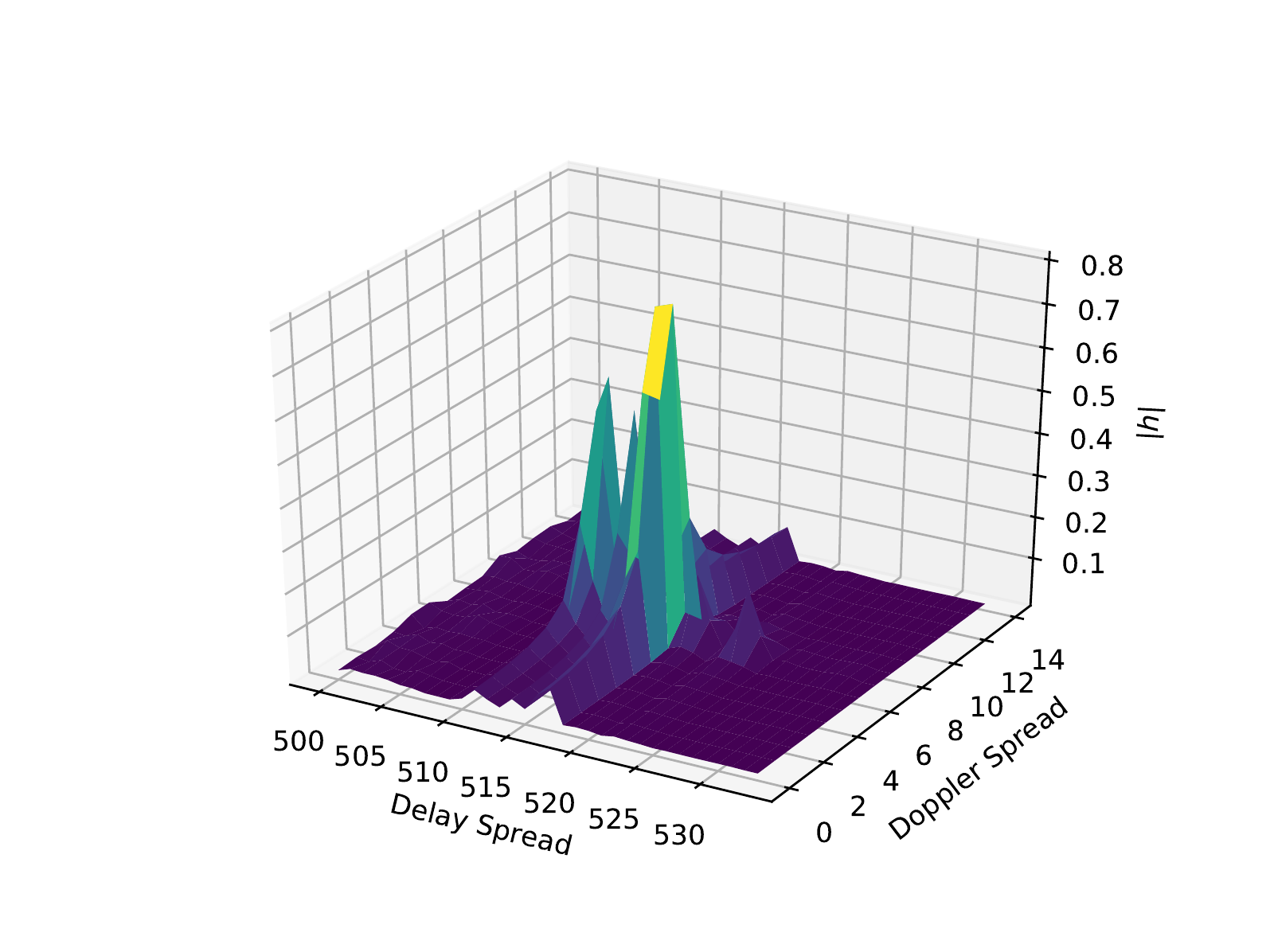}
		\label{dd}}
    \caption{Channel representations in different domain:  (a) in the time-frequency domain, (b) in the delay-Doppler domain.}
    \label{channel_delay_doppler}
\end{figure}

We consider a wireless communication system with a transmitter and a receiver. Given the time domain transmitted signal $x(t)$, the received signal $r(t)$ is given by
\begin{equation}
\label{Rx_signal}
\begin{aligned}
r(t)&= \int {\tilde h}(\tau, t) x(t-\tau) d \tau \\
\end{aligned},
\end{equation}
where $\tau$ denotes delay and $t$ denotes time in the doubly selective channel response ${\tilde h}(\tau, t)$. Using ${\tilde h}(\tau, t)$, the frequency selectivity of the wireless channel is characterized by having multi-path coefficients on $\tau$. The time selectivity of the wireless channel is shown as ${\tilde h}(\tau, t_1) \neq {\tilde h}(\tau, t_2)$ when $t_1 \neq t_2$. Alternatively, the input-output relation in (\ref{Rx_signal}) can be expressed via a Doppler-variant channel impulse response as follows,
\begin{equation}
\begin{aligned}
    r(t) = \iint h(\tau, \nu) x(t-\tau) e^{j 2 \pi \nu t} d \tau d \nu
\end{aligned}
\end{equation}
where 
\begin{equation*}
\begin{aligned}
 h(\tau, \nu) := \int {\tilde h}(\tau, t)  e^{- j 2 \pi \nu t} dt,
\end{aligned}
\end{equation*}
is the so-called Doppler-variant impulse response. 
To be specific, it represents the Doubly-selective channel via the delay-Doppler domain. 
Note that the above formulation has a missing internal term $e^{-2\pi\nu\tau}$ compared to equation (\ref{Rx_signal}) in \cite{hadani2017orthogonal}. This is because the missing term is part of $h(\tau,\nu)$ in our notation. Representations of a doubly selective channel on these two domains are plotted in Fig.~\ref{channel_delay_doppler}.

\subsection{OTFS Modulation}
\begin{figure}
    \centering
    \includegraphics[width = 1\linewidth]{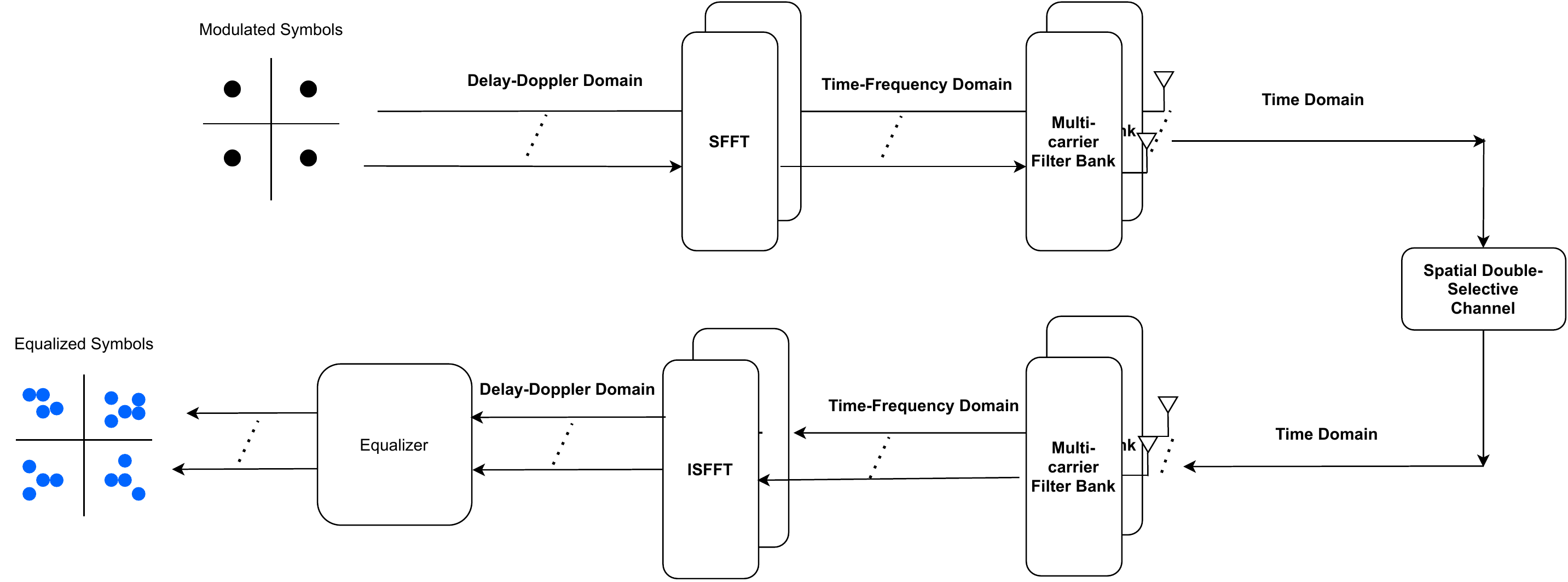}
    \caption{A link level schematic of a MIMO-OTFS system.}
    \label{RC_OTFS}
\end{figure}

An implementation of OTFS modulation is shown in Fig. \ref{RC_OTFS}, where OTFS is realized by adding pre-and post-processing blocks to a standard multi-carrier modulation system. The pre-processing block converts modulation symbols from the delay-Doppler domain to the time-frequency domain and vice versa.
The multi-carrier filter bank (Heisenberg transform) at the Tx converts a 2D time-frequency domain signal to a 1D time-domain waveform. In our notation, we define the time domain waveform of one {\textbf{OTFS frame}} as $x(t)$. It is written as,
\begin{equation}
\label{time_seq}
x(t)=\sum_{n=0}^{N-1} \sum_{m=0}^{M-1} X[n, m] g_{\mathrm{tx}}(t-n \Delta T) e^{j 2 \pi m \Delta f(t-n \Delta T)},
\end{equation}
where $X[n,m]$ is its time-frequency representation, $g_{tx}(t)$ is the pulse shaping filter, $M$ stands for the number of frequency tones, $\Delta T$ is the time duration of one transmitting pulse, $\Delta f$ is the frequency spacing, and $N$ is the number of transmitted pulses in one OTFS frame. Therefore, an OTFS frame occupies $N\Delta T$ seconds in time and $M\Delta f$ Hz bandwidth. 

The pre-processing block which transforms modulation symbols from the delay-Doppler domain to the time-Frequency domain is defined as,
\begin{equation}
\label{ISFFT}
X[n, m]=\frac{1}{\sqrt{N M}} \sum_{k=0}^{N-1} \sum_{l=0}^{M-1} x[k, l] e^{j 2 \pi\left(\frac{n k}{N}-\frac{m l}{M}\right)}.
\end{equation}
where $x[k,l]$ are the modulation symbols residing in the delay–Doppler domain. The transformation (\ref{ISFFT}) is also referred to as the inverse symplectic finite Fourier transform (ISFFT) since it has Fourier basis functions. In addition, $x[n,m]$ can be interpreted as the superposition of QAM symbols spreading over the full time-frequency grid using $e^{j 2 \pi\left(\frac{n k}{N}-\frac{m l}{M}\right)}$ as the basis functions. Therefore, OTFS is able to exploit the diversity across the full-time and frequency domain. At the receiver, the received signals at the time-frequency domain, and delay-Doppler domain are respectively denoted as $Y[n,m]$ and $y[k,l]$. The relation between $Y[n,m]$ and $y[k,l]$ is characterized by the SFFT. $Y[n,m]$ is obtained by using the receiving filter bank, which has a similar formulation (\ref{time_seq}). For simplicity, we summarized the notations of the OTFS system in Table~\ref{OTFS_notations}.

\subsection{Integration with OFDM Systems}
\label{c_ofdm}
\begin{figure}
    \centering
    \includegraphics[width = 1\linewidth]{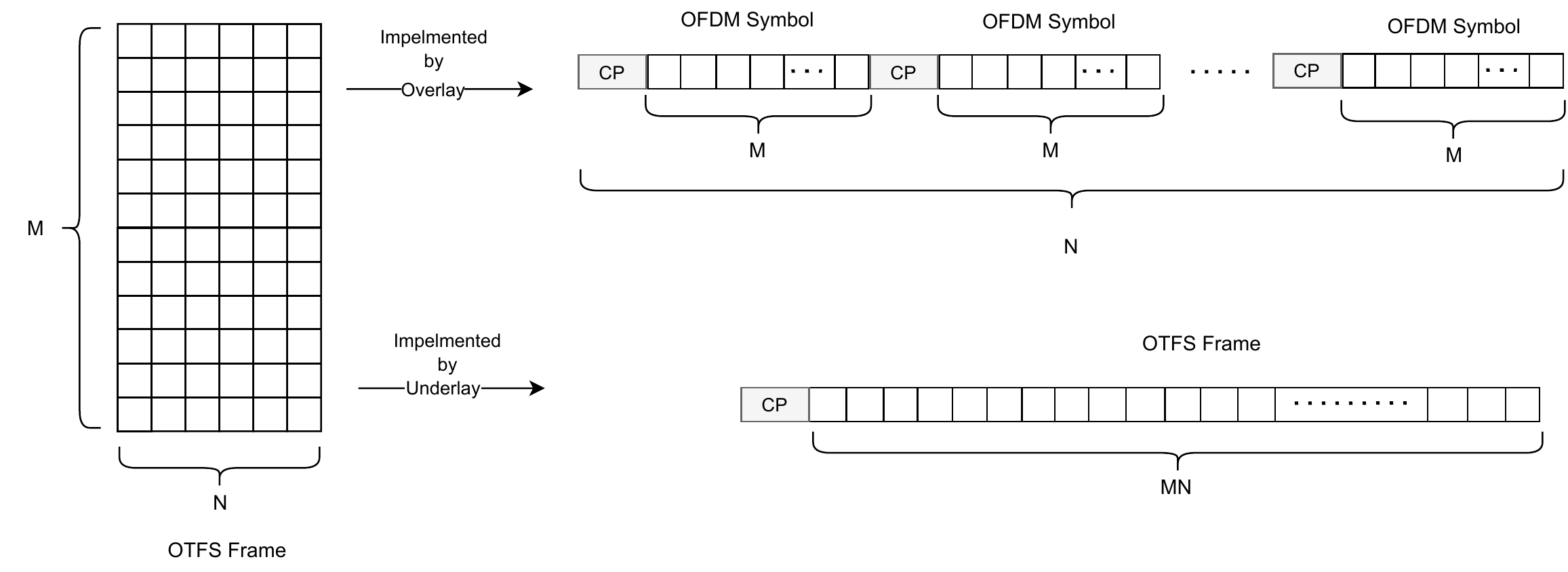}
    \caption{The time-domain frame structure of two different OTFS implementations: the top one overlays an OFDM system, and the bottom one is a standalone version.}
    \label{implementation_OTFS}
\end{figure}

As shown in Fig. \ref{RC_OTFS}, an OTFS system can be implemented as an overlay of an OFDM system. This implementation expedites prototyping since OFDM systems are highly optimized in current wireless systems, such as 5G NR and WiFi. However, building an OTFS system on top of an existing transceiver system brings extra complexity. Alternatively, an OTFS system can be realized via its own ``standalone''. This simplification is because half of the operations on the pre-and post-processing operations (i.e., ISFFT and SFFT) can be canceled out by the FFT and IFFT operations from the multi-carrier filter bank. Therefore, the resulting standalone implementation is more lightweight, since it only needs one pair of FFT and IFFT in the transceiver chain. Meanwhile, the cyclic prefix (CP) of each OFDM symbol can be eliminated in OTFS, significantly improving spectral efficiency. Without adding CP inside the OTFS frame, OTFS handles the inter-symbol interference in a frame-based manner, where the interference ``between OFDM symbols'' is lumped to a 2D basis to equalize. However, a CP still needs to be added between two consecutive OTFS frames to avoid the ``frame interference''. As a summary, Fig. \ref{implementation_OTFS} depicts the time frame structures of these two different schemes for OTFS system implementation under the same sampling rate (an overlay on an OFDM system and a standalone version). 

\subsection{End-to-End Channel Model in the Delay-Doppler Domain}
For simplicity, we assume ideal bi-orthogonal transmit and receive pulse shaping filters\cite{hadani2017orthogonal}. After substituting (\ref{time_seq}) and (\ref{ISFFT}), as well as their counterparts at the receiver, into (\ref{Rx_signal}), we arrive at an end-to-end model connecting $x[k,l]$ and $y[k,l]$ within one OTFS frame. The relation is given below \cite{hadani2018otfs}, 
\begin{equation}
\begin{aligned}
y[k, l]&=\frac{1}{N M} \sum_{k^{\prime}=0}^{N-1} \sum_{l^{\prime}=0}^{M-1} {\bar x}\left[k^{\prime}, l^{\prime}\right] h\left[k-k^{\prime}, l-l^{\prime}\right]\\
k &= 0, \cdots, N-1; 
l = 0, \cdots, M-1
\end{aligned}
\end{equation}
where ${\bar x}[k', l']$ is the periodized version of $x[k', l']$ with periods $(1/{\Delta T}, 1/{\Delta f})$ in the delay-Doppler domain, and 
\begin{equation}
h\left[k^{\prime}, l^{\prime}\right] :=\left.h(\nu, \tau)\right|_{\nu=\frac{k^{\prime}}{N \Delta T}, \tau=\frac{l^{\prime}}{M \Delta f}}.
\end{equation}
Note that CP added between two consecutive OTFS frames accomplishes this 2D circular convolution relation. In addition, any imperfections in transmit and receive pulse shaping can be absorbed into the channel kernel $h[k, l]$ for a more comprehensive formulation. 

\begin{figure}
    \centering
	\subfloat[]{\includegraphics[width=0.75\linewidth]{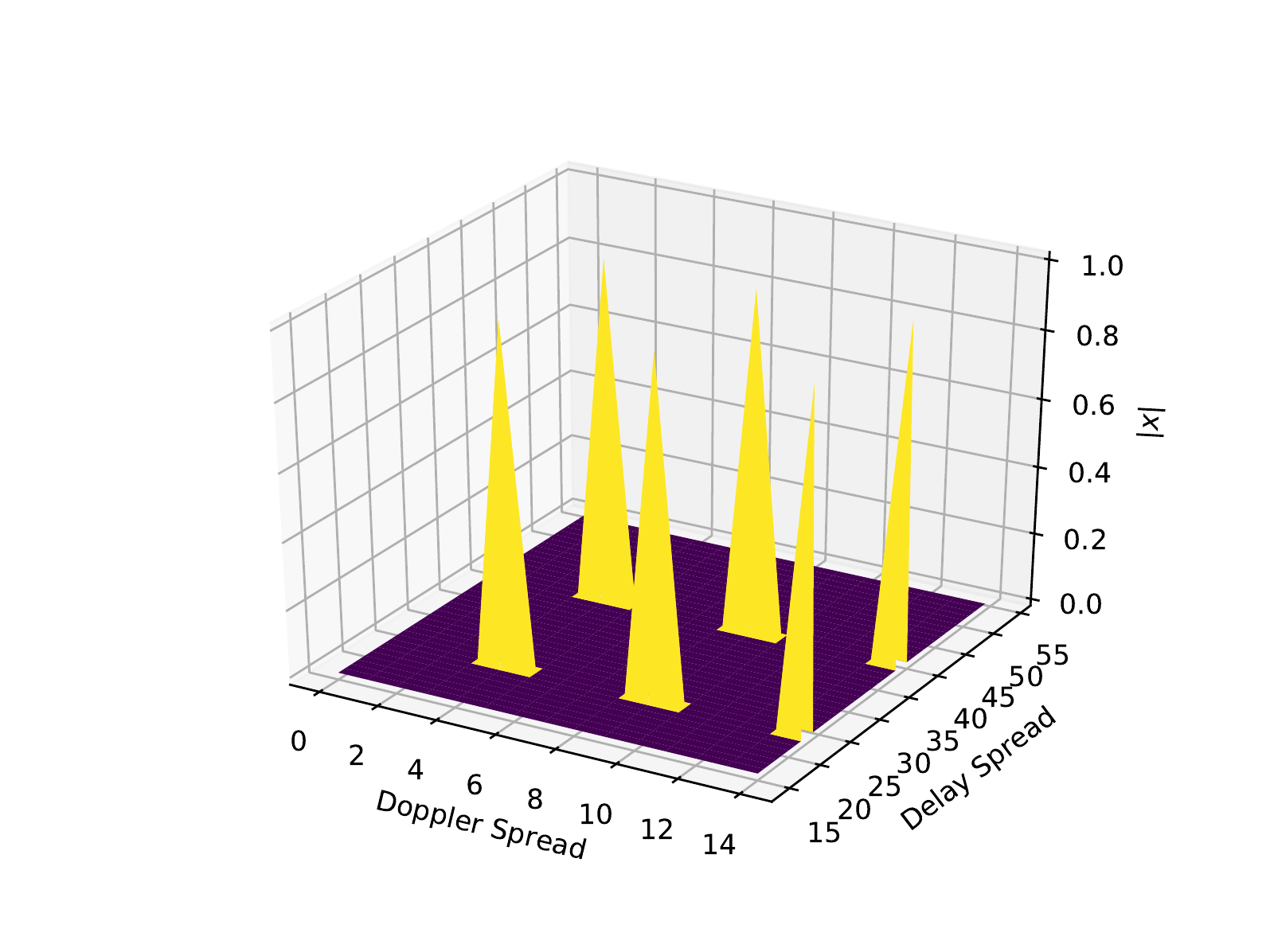}
		\label{tf}}
	\vfil
	\subfloat[]{\includegraphics[width=0.75\linewidth]{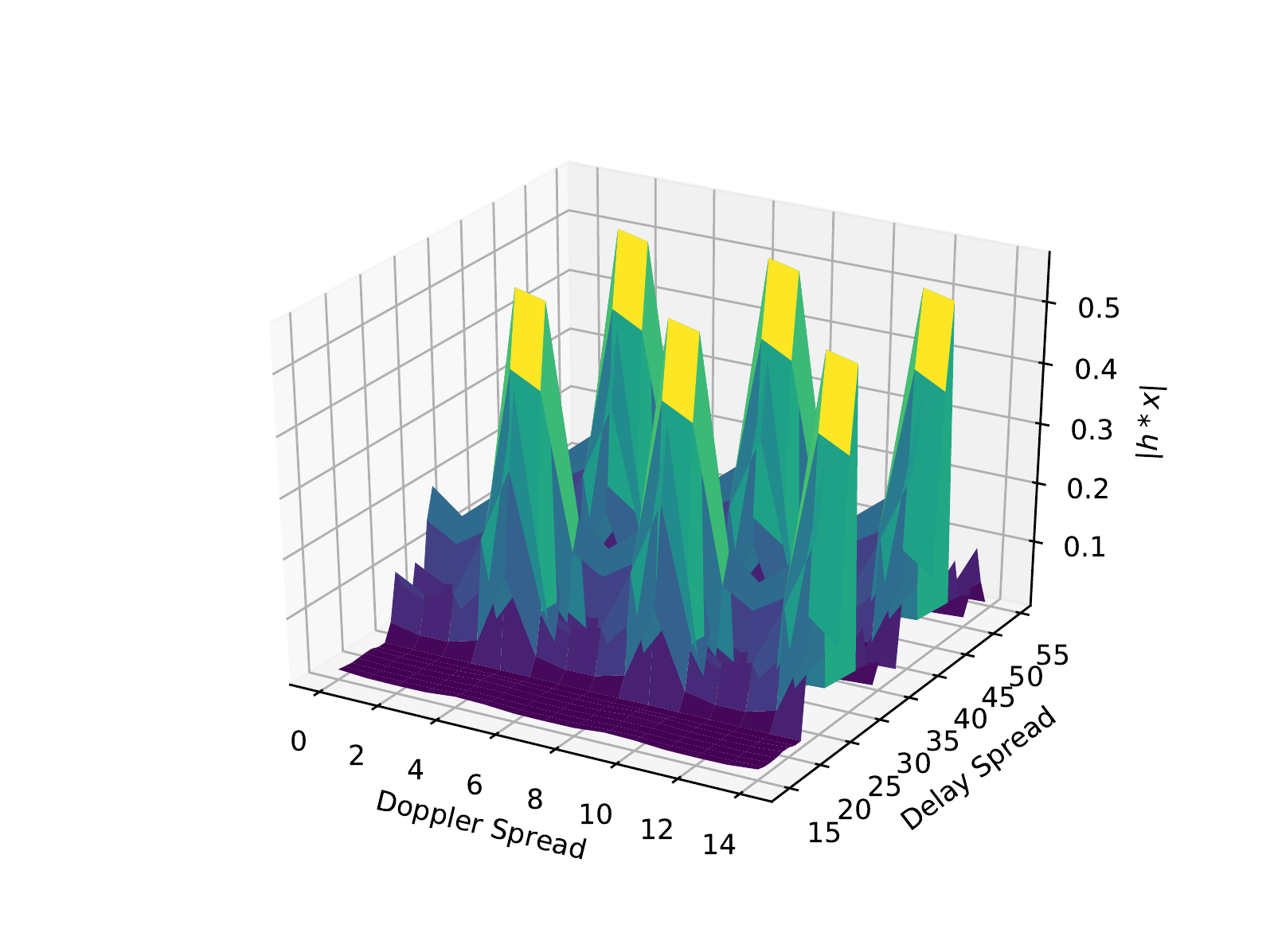}
		\label{dd}}
    \caption{Distribution of symbols in the delay-Doppler domain at the transmitter (a) and the receiver (b).}
    \label{input_output_relation}
\end{figure}


\begin{table}[h]
\centering
\caption{Notation in OTFS Systems}
\begin{tabular}{|l|l|l|}
\hline
Notations & Definitions \\ \hline
$x[k,l]$ & Modulation Symbols in Delay-Doppler Domain at Tx \\ \hline
$X[n,m]$ & Time-Frequency Domain representation of the modulated symbols (via ISFFT) at Tx \\ \hline
$x(t)$ & Time Domain Waveform of OTFS Signal at Tx \\\hline
${\tilde h}(\tau, t)$ & Doubly-Selective Channel in delay-time domain \\ \hline
${h}(\tau, \nu)$ & Doubly-Selective Channel in delay-Doppler domain \\ \hline
$r(t)$ & Time Domain Waveform of OTFS Signal at Rx \\\hline
$N$ & Number of transmitted pulses in one OTFS frame\\ \hline
$M$ & Number of sub-carriers in one OTFS frame \\ \hline
${y}[k,l]$ & Modulation Symbols in Delay-Doppler Domain at Rx\\ \hline
${Y}[n,m]$ & Time-Frequency Domain representation of the modulated symbols (via ISFFT) at Rx\\ \hline
\end{tabular}
\label{OTFS_notations}
\end{table}

\section{RC Meets OTFS in Equalization}
\label{learning_method}
We first introduce two types of online training dataset in an OTFS system with single antennas at the transmitter and receiver. We then present a neural network structure and the associated learning algorithms for equalization at the receiver.

\subsection{Training and Testing Dataset}
\label{pilot}
Since one OTFS frame can be discretized as a grid with $N$ intervals along the Doppler domain and $M$ intervals along the delay domain, we denote the modulated symbols $x[k, l]$ over one OTFS frame as a matrix ${\boldsymbol X} \in {\mathcal C}^{M \times N}$, where $\mathcal C$ is the predefined constellation for digital modulation, such as ${\mathcal C} = \{+1, -1\}$ for BPSK, and ${\mathcal C}  = \{+1+1j, +1-1j, -1+1j, -1-1j \}$ for QPSK. We define the online training dataset (i.e., pilot symbols) and testing dataset within the same OTFS frame. Accordingly, the equalization operation is based on a single OTFS frame: The training method only uses the pilot symbols from one OTFS frame; Then, it applies the learned neural network to equalize the data symbols in the remainder of the same OTFS frame. This operation mode is robust to sudden channel changes, such as when the base station adopts a new transmission rank and schedules a new user. This paper considers two ways to construct pilots in the delay-Doppler domain: interleaved pilot and superimposed pilot patterns.

\subsubsection{Interleaved Pilot}
\begin{figure}[h]
    \centering
    \includegraphics[width = 0.85\linewidth]{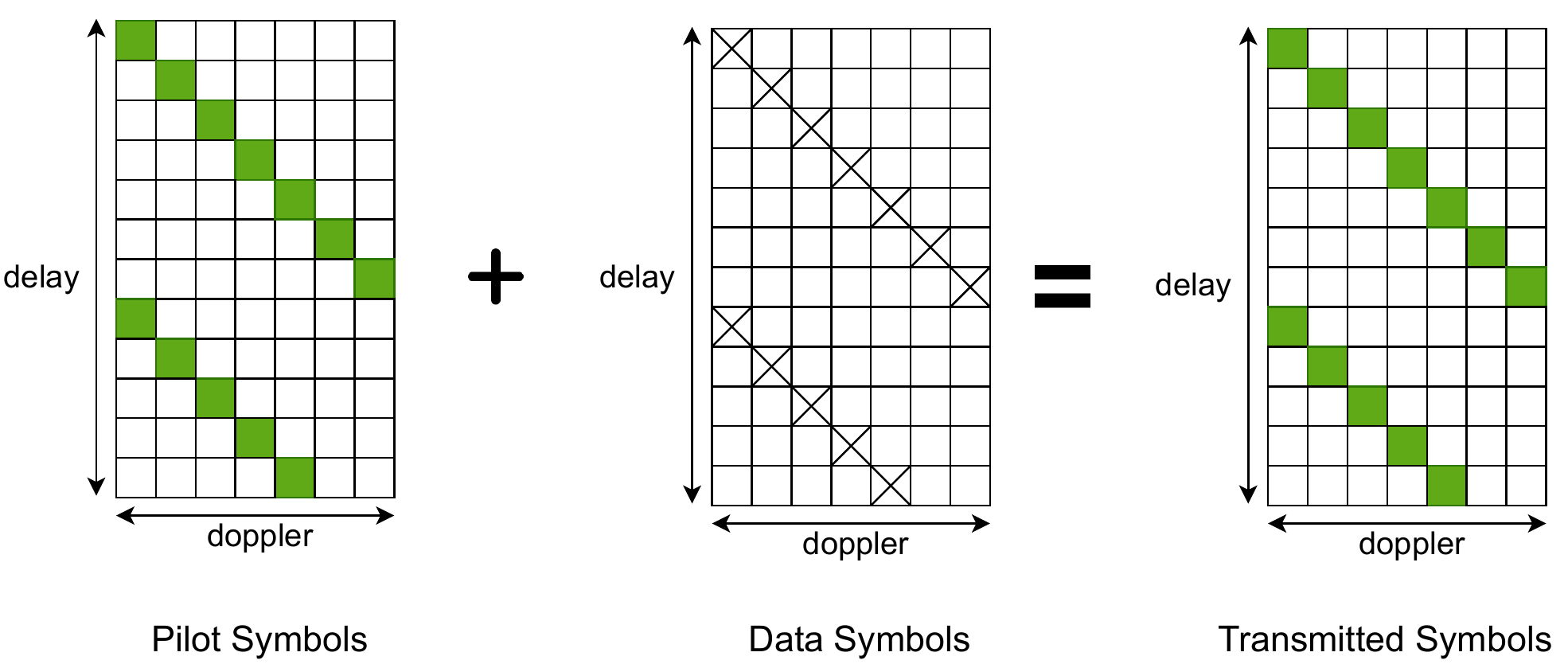}
    \caption{Delay-Doppler representation of interleaved pilots associated with the data symbols in one OTFS frame.}
    \label{pilot_design}
\end{figure}

\begin{figure}[h]
    \centering
    \includegraphics[width = 0.85\linewidth]{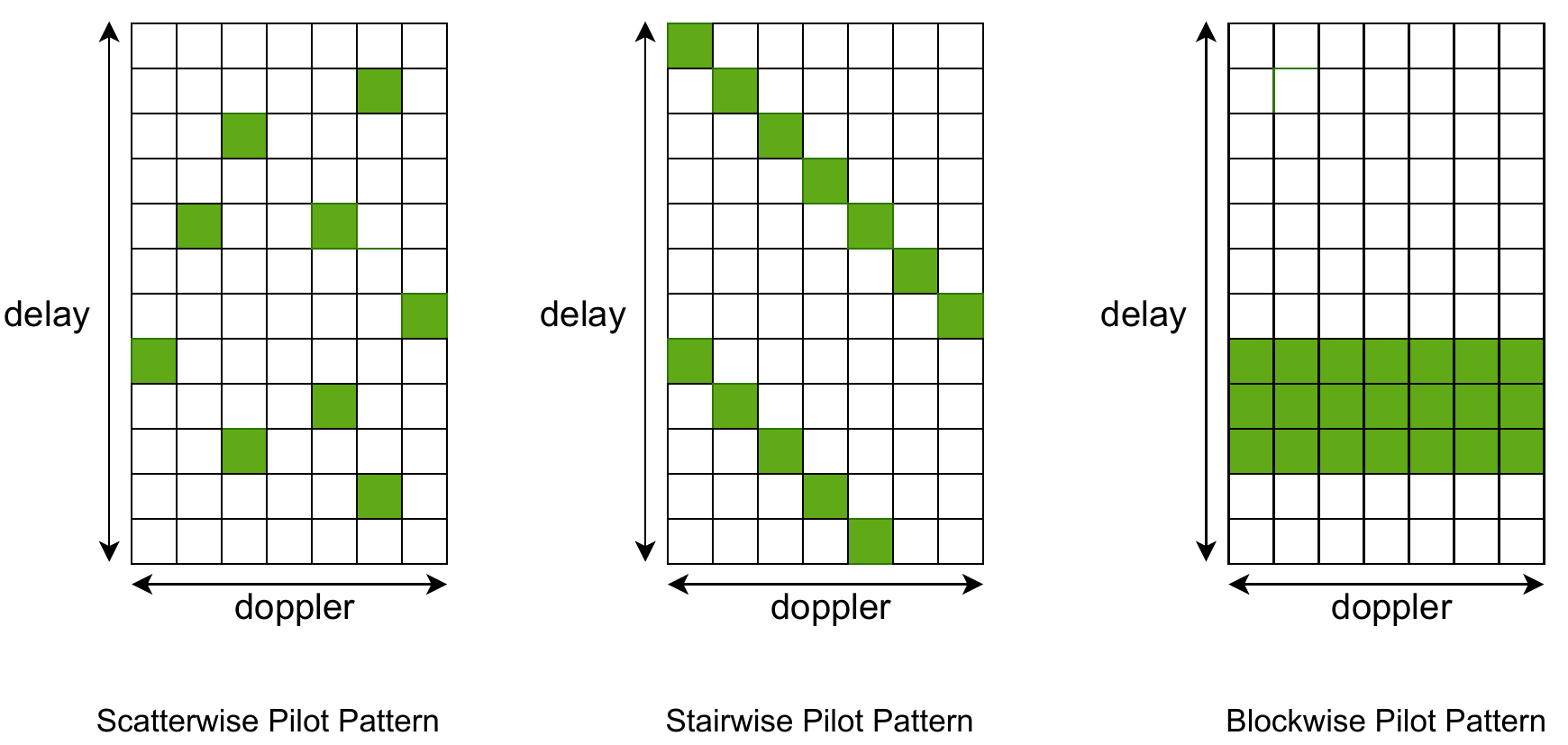}
    \caption{Different interleaved pilot patterns over one OTFS frame represented in the delay-Doppler Domain.}
    \label{pilot_design2}
\end{figure}

Here pilot symbols interleave with data symbols in the delay-Doppler domain as shown in Fig. \ref{pilot_design}. The training dataset is denoted as
\begin{align}
    \label{interleaved_training_set}
    \{{\boldsymbol X}_{train} := {\boldsymbol \Omega}\odot {\boldsymbol X}, {\boldsymbol Y}_{train} := {\boldsymbol Y}\},
\end{align}
where \textit{inside any of the curly brackets above and below, the left-hand element represents the NN desired output, while the right-hand component stands for the NN input,} and $\odot$ stands for the Hadamard product between two matrices. Accordingly, the testing dataset is given by, 
\begin{align}
    \{{\boldsymbol X}_{test} := {\bar{\boldsymbol \Omega}}\odot {\boldsymbol X}, {\boldsymbol Y}_{test} := {\boldsymbol Y}\},
\end{align}
where $\boldsymbol X$ is a symbol matrix representing one OTFS frame at the transmitter, ${\boldsymbol \Omega}$ is an indication matrix with value $1$ at the location assigned as pilot symbols, and value $0$ elsewhere, and $\boldsymbol Y$ is the corresponding received OTFS frame at the receiver. Fig. \ref{pilot_design} only shows one possible realization of the interleaved pilot pattern, where the stairwise green spots in the first sub-figure stand for pilots (training dataset), and the remaining blank blocks represent data symbols for transmission (testing dataset). Alternative groupings of pilot symbols are shown in Fig. \ref{pilot_design2}. The pilot overhead $\eta$ for the interleaved pilot is given by,
\begin{align}
    \label{overhead}
    \eta = \frac{|{\boldsymbol \Omega}|}{M\cdot N}.
\end{align} 
Note that both the training and testing dataset use the observation $\boldsymbol Y$, rather than ${\boldsymbol Y}\odot{\boldsymbol \Omega}$ and ${\boldsymbol Y}\odot{\bar{\boldsymbol \Omega}}$ respectively. This design is motivated by the objective of learning the mutual interference between pilot and symbols to further reduce training overhead. It is possible to add guard symbols between pilot and data symbols, and this approach has been adopted in conventional channel estimation frameworks for OTFS, such as \cite{raviteja2019embedded}\footnote{The guard symbols defined in \cite{raviteja2019embedded} can be absorbed into the support of $\Omega$ when calculating the pilot overhead.}. 

\subsubsection{Superimposed Pilot}
\begin{figure}
    \centering
    \includegraphics[width = 0.85\linewidth]{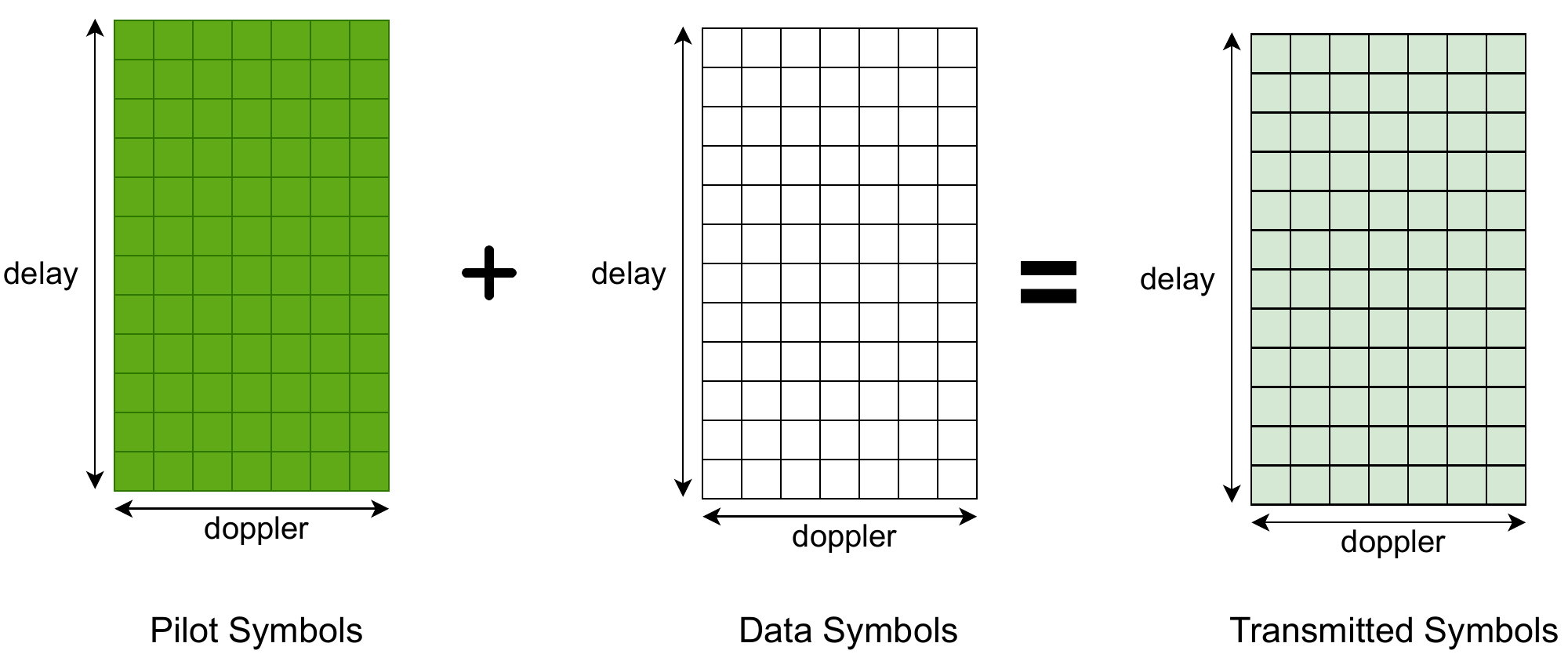}
    \caption{Superimposed Pilot in the delay-Doppler Domain}
    \label{pilot_design3}
\end{figure}

Here pilot symbols are spread over the 2D delay-Doppler domain, and are superimposed on data symbols as illustrated in Fig. \ref{pilot_design3}. Therefore, symbols in each transmitted OTFS frame can be written as ${\boldsymbol X} = {\boldsymbol X}_{train} + {\boldsymbol X}_{test}$, where ${\boldsymbol X}_{test} \in {\mathcal C}^{M \times N}$, and ${\boldsymbol X}_{train} \in {\mathbb C}^{M \times N}$. Since the ideal end-to-end relation in OTFS system can be formulated as a twisted convolution (a linear mapping) on the 2D plane, the received OTFS symbols are given by ${\boldsymbol Y} = {\boldsymbol Y}_{train} + {\boldsymbol Y}_{test}$, where ${\boldsymbol Y}_{train}$ corresponds to the receiving components contributed only from ${\boldsymbol X}_{train}$, and ${\boldsymbol Y}_{test}$ is from ${\boldsymbol X}_{test}$. 

However, simply adding superimposed pilots over data symbols may not be easy for the receiving process. Regarding a supervised learning-based equalization framework, the receiver has to first identify the pilot components from received symbols $\boldsymbol Y$. Rather than using the support of interleaved pilots as discussed in the previous subsection, we consider resorting to the time-frequency domain to distinguish pilot symbols and data symbols. To this end, we design our superimposed pilot symbols as follows,
\begin{align*}
    {\boldsymbol X}_{train} = {\text{SFFT}}(c\cdot{\boldsymbol \Omega}),
\end{align*}
${\text{SFFT}}$ represents the inverse mapping of (\ref{ISFFT}) operating on matrices, and $c$ is a constant which is expected to suppress the value of any entries of ${\text{ISFFT}}({\boldsymbol X}_{test})$ to make pilots distinguishable from data symbols. 

However, a large constant $c$ leads to a high peak to average ratio (PAPR) of the transmitted signal. We can add an interference term to diminish the energy from data symbols at the support of ${\boldsymbol \Omega}$. To this end, we incorporate a helper interference term defined as follows,
\begin{align*}
    {\boldsymbol X}_{aid-int} = -{\text{SFFT}}({\text{ISFFT}}({\boldsymbol X}_{test})\odot{\boldsymbol \Omega}).
\end{align*}
The newly formed transmitted signals in the delay-Doppler domain are,
\begin{equation}
\begin{aligned}
    \label{sup_pilot_rx_signal}
    {\boldsymbol X} = {\boldsymbol X}_{train} + {\boldsymbol X}_{test} + {\boldsymbol X}_{aid-int},
\end{aligned}
\end{equation}
and the corresponding received signals are $\boldsymbol Y$. Therefore, the training dataset has the following compact expression, 
\begin{align}
        \{{\boldsymbol X}_{train} := {\text{SFFT}}( {\boldsymbol \Omega}), {\boldsymbol Y}_{train} := {\text{SFFT}}({\text{ISFFT}}({\boldsymbol Y})\odot {\boldsymbol \Omega})\},
\end{align}
The testing dataset is,
\begin{align}
        \{{\boldsymbol X}_{test} , {\boldsymbol Y}_{test} := {\boldsymbol Y}\}.
\end{align}

Note that the overhead calculation of the superimposed pilot is not straightforward. If we were to directly calculate pilot overhead based in the delay-Doppler domain, the pilot overhead would likely be $0\%$ since data symbols spread over the entire grid in the delay-Doppler domain. 
However, we may calculate overhead in terms of power required by the superimposed pilots. Given the duality between time-frequency and delay-Doppler, we may still calculate the overhead by using (\ref{overhead}) in the time-frequency domain.



\subsection{Neural Network Structure for Equalization -- Reservoir Computing}
Reservoir computing (RC) is a memory-enabled neural network framework. 
A vanilla version of RC comprises a nonlinear activation function, an input layer, a recurrent layer, and an output layer. 
The computational power of RC is realized by mapping the recurrent states to desired signals through learning the output layers. 
Furthermore, due to the fixed reservoir dynamics, RC training is only conducted on the readout layers, which effectively avoids gradient vanishing/explosion issues by avoiding back-propagation through time.

A discrete-time realization of RC can be characterized by its internal state transition and output mapping. The state transition is formulated as follows,
\begin{align}
\label{rec_layer}
{\boldsymbol s}(t+1) = \sigma\left({\boldsymbol W}_{tran}
\begin{bmatrix}
{\boldsymbol s}(t)\\
{\boldsymbol y}(t)
\end{bmatrix}\right)
\end{align}
where $t$ stands for the time index, $\sigma$ is a nonlinear function, ${\boldsymbol s}(t)$ is a vector representing the internal reservoir state, ${\boldsymbol y}(t)$ is the input vector, and ${\boldsymbol W}_{tran}$ stands for the reservoir weight matrix which is often chosen with a spectral radius smaller than 1 in order to asymptotically achieve a ``similarity'' to the desired dynamic model\cite{jaeger2001echo}. The output equation is simply treated as
\begin{align}
\label{out_layer}
{\boldsymbol z}(t) = {\boldsymbol W}_{out}
\begin{bmatrix}
{\boldsymbol s}(t)\\
{\boldsymbol y}(t)
\end{bmatrix},
\end{align}
where ${\boldsymbol W}_{out}$ is the output weight matrix and ${\boldsymbol z}(t)$ stands for the output. We can stack the features in an extended state matrix 
\begin{align}
\label{extend_state}
\bar {\boldsymbol S} = \left[ 
\begin{bmatrix*}
{\boldsymbol s}(0)\\
{\boldsymbol y}(0)
\end{bmatrix*},
\begin{bmatrix*}
{\boldsymbol s}(1)\\
{\boldsymbol y}(1)
\end{bmatrix*},
\cdots,
\begin{bmatrix*}
{\boldsymbol s}(t)\\
{\boldsymbol y}(t)
\end{bmatrix*}, \right]^T.
\end{align}
The loss function for the output layer learning is then given by, 
\begin{align}
    \min_{\boldsymbol W} \| {\mathcal RC}_{\boldsymbol W}({\boldsymbol Y}_{train}) - {\boldsymbol X}_{train}\|_F
\end{align}
where the input and output dimensions of ${\mathcal RC}_{\boldsymbol W}(\cdot)$ are configured as $N$, and the sequence length is set as $M$. Once ${\boldsymbol Y}_{train}$ is fed into RC, an internal extended state matrix $\bar {\boldsymbol S}$ can be obtained ready for output layer learning as suggested by (\ref{out_layer}).

\subsection{Learning Algorithms}
We introduce two different ways to learn the output weights of RC in the following two subsections corresponding to the two different pilot patterns. 

\subsubsection{Interleaved Pilot}
Here we define the loss function as 
\begin{align*}
    \min_{\boldsymbol W}\|{\boldsymbol X}_{train} - {\boldsymbol \Omega}\odot({\boldsymbol {\bar S}}_{train}{\boldsymbol W}_{out})\|_{F}^2.
\end{align*}
A closed-form solution of the above problem can be obtained via rephrasing the objective function as,
\begin{align*}
    \min_{\boldsymbol W} \sum_{n = 1}^N \|(({\boldsymbol X}_{train})_{:,n}- {\text{diag}}({\boldsymbol \Omega}_{:,n}){\boldsymbol {\bar S}}_{train}({\boldsymbol W}_{out})_{:,n}]) \|_2^2
\end{align*}
By solving the above separable least squares, the closed-form solution of ${\boldsymbol W}_{out}$ is given by each of its column vectors as follows,
\begin{align}
\label{readout_scatter}
\left({\boldsymbol W}_{out}\right)_{:, n} = \left[{\text{diag}}\left({\boldsymbol \Omega}_{:,n}\right){\boldsymbol {\bar S}}_{train}\right]^+\left({\boldsymbol X}_{train}\right)_{:, n}.
\end{align}
The transmitted data symbols are estimated by applying the output weights through, 
\begin{align}
\label{sd_scatter}
{\boldsymbol {\hat X}}_{test} = {\mathcal Q}\left({\boldsymbol {\bar S}}_{test}{\boldsymbol W}_{out}\right)\odot{\boldsymbol \Omega}.
\end{align}
where $\mathcal Q$ is the optimal quantizer for the constellation of $\mathcal C$. Note that ${\bar{\boldsymbol S}_{train}} = {\bar{\boldsymbol S}}_{test}$ since ${\boldsymbol Y}_{train} = {\boldsymbol Y}_{test}$.

\subsubsection{Superimposed Pilot}

The objective of superimposed pilot based training is given as follows,
\begin{align*}
    \min_{\boldsymbol W}\|{\boldsymbol X}_{train} - {\boldsymbol {\bar S}}_{train}{\boldsymbol W}_{out}\|_{F}^2,
\end{align*}
We can directly obtain the coefficients of the readout weight as follows,
\begin{align}
    \label{readout_superimposed}
    {\boldsymbol W}_{out} = {\bar{\boldsymbol S}_{train}}^+{\boldsymbol X}_{train}.
\end{align}
Then the transmitted data symbols are obtained  by
\begin{align}
    \label{sd_superimpose}
    {\boldsymbol {\hat X}}_{test} = {\mathcal Q}({\boldsymbol {\bar S}_{test}}{\boldsymbol W}_{out}).
\end{align}
Note that quantization from RC output to the estimated symbols essentially eliminates the contribution from the helper interference term in (\ref{sup_pilot_rx_signal}). The learning and testing stages for the above two different pilot patterns are summarized jointly in Algorithm \ref{algorithm0}. 



\begin{algorithm}[!ht]
	\caption{RC for OTFS Equalization}
	\label{algorithm0}
	\begin{algorithmic}
		\REQUIRE $\{{\boldsymbol Y}_{train}, {\boldsymbol X}_{train}\}$, ${\boldsymbol Y}_{test}$
		\ENSURE ${\boldsymbol W}_{out}$, $\hat{\boldsymbol X}_{test}$
		\FOR{$l \in [0, L_{forget}]$}
		\STATE Input sequence $[{\boldsymbol Y}, {\boldsymbol O}_l]$ into RC and obtain a delayed extended state matrix ${\bar{\boldsymbol S}}_l := {\bar{\boldsymbol S}}(:,l:)$, where $\bar{\boldsymbol S}$ is the extended state matrix as defined in (\ref{extend_state}), and ${\boldsymbol O}_l$ is a zero matrix with $l$ columns.
        \STATE Calculate a corresponding output layer ${\boldsymbol W}_{out}^l$ by using (\ref{readout_scatter}) for interleaved pilots and (\ref{readout_superimposed}) for superimposed pilots.
		\STATE Cache the ${\text {LOSS}}_l$ value of the learned output weight.
		\ENDFOR{}
		\STATE Sort the optimal $l^{\star} =\arg\min_p {\text{LOSS}}_p$, and select ${\boldsymbol W}_{out} = {\boldsymbol W}_{tout}^{l^{\star}}$
		\STATE Obtain $\hat{\boldsymbol X}_{test}$ by using (\ref{sd_scatter}) for interleaved pilot and (\ref{sd_superimpose}) for superimposed pilot.
	\end{algorithmic}
\end{algorithm}

\section{Extensions and Enhancements}
\label{Ext}
\subsection{Learning in MIMO-OTFS}
We assume there are $N_t$ and $N_r$ antennas at transmitter and receiver, respectively. The end-to-end relation between transmitted and received symbols is given by,
\begin{equation}
\begin{aligned}
y_{n_r}[k, l]&=\frac{1}{N M} \sum_{n_t = 0}^{N_t -1} \sum_{k^{\prime}=0}^{N-1} \sum_{l^{\prime}=0}^{M-1} {\bar x}_{n_t}\left[k^{\prime}, l^{\prime}\right] h_{n_r, n_t}\left[k-k^{\prime}, l-l^{\prime}\right]\\
k &= 0, \cdots, N-1; 
l = 0, \cdots, M-1
\end{aligned},
\end{equation}
where $h_{n_r, n_t}[k, l]$ denotes the channel between the $n_r$-th received antenna and the $n_t$-th transmitted antenna, represented in the delay-Doppler domain, and ${\bar x}_{n_t}$ is the periodized representation of the transmitted symbols at the $n_t$th antenna in the delay-Doppler domain. 

In the MIMO scenario, we suppose the pilot pattern at any single Tx antennas of the MIMO system follows the same design as the SISO scenario; at the same time, it is reused by different Tx antennas. This pilot structure allows the neural network to perceive the interference pattern of the spatial channel at the receiver side. By contrast, conventional channel estimation methods often employ non-overlapping pilots to avoid antenna interference. To process the RC based learning, we stack the transmitted symbols as a three-mode matrix ${\boldsymbol {\mathcal X}}\in {\mathcal C}^{M \times N \times N_t}$. Accordingly, the training dataset is generalized as follows
\begin{align*}
\left\{{\boldsymbol X}_{train} := 
    \begin{bmatrix}
    {\boldsymbol \Omega}\\
    {\boldsymbol \Omega}\\
    \vdots \\
    {\boldsymbol \Omega}
    \end{bmatrix}    
    \odot
    \begin{bmatrix}
    {\boldsymbol {\mathcal X}}_{:,:,1}\\
    {\boldsymbol {\mathcal X}}_{:,:,2}\\
    \vdots \\
    {\boldsymbol {\mathcal X}}_{:,:,N_t}
    \end{bmatrix}
    , {\boldsymbol Y}_{train} := \begin{bmatrix}
    {\boldsymbol {\mathcal Y}}_{:,:,1}\\
    {\boldsymbol {\mathcal Y}}_{:,:,2}\\
    \vdots \\
    {\boldsymbol {\mathcal Y}}_{:,:,N_t}
    \end{bmatrix}\right\}.
\end{align*}
and the testing dataset is given by
\begin{align*}
\left\{{\boldsymbol X}_{test} :=     
\begin{bmatrix}
    \bar{\boldsymbol \Omega}\\
    \bar{\boldsymbol \Omega}\\
    \vdots \\
    \bar{\boldsymbol \Omega}
    \end{bmatrix}    
    \odot
    \begin{bmatrix}
    {\boldsymbol {\mathcal X}}_{:,:,1}\\
    {\boldsymbol {\mathcal X}}_{:,:,2}\\
    \vdots \\
    {\boldsymbol {\mathcal X}}_{:,:,N_t}
    \end{bmatrix}, {\boldsymbol Y}_{test} := \begin{bmatrix}
    {\boldsymbol {\mathcal Y}}_{:,:,1}\\
    {\boldsymbol {\mathcal Y}}_{:,:,2}\\
    \vdots \\
    {\boldsymbol {\mathcal Y}}_{:,:,N_t}
    \end{bmatrix}\right\}.
\end{align*}

Similarly, the training and testing dataset for the superimposed pilot in the MIMO system are,
\begin{align}
\left\{{\boldsymbol X}_{train} := 
\begin{bmatrix}
{\text{SFFT}}({\boldsymbol \Omega})\\
        {\text{SFFT}}({\boldsymbol \Omega})\\
        \vdots \\
        {\text{SFFT}}({\boldsymbol \Omega})
        \end{bmatrix}
        , {\boldsymbol Y}_{train} :=
        \begin{bmatrix}
        {\text{SFFT}}({\text{ISFFT}}({\boldsymbol {\mathcal Y}}_{:,:,1})\odot {\boldsymbol \Omega}) \\
        {\text{SFFT}}({\text{ISFFT}}({\boldsymbol {\mathcal Y}}_{:,:,2})\odot {\boldsymbol \Omega})\\
        \vdots \\
        {\text{SFFT}}({\text{ISFFT}}({\boldsymbol {\mathcal Y}}_{:,:,N_t})\odot {\boldsymbol \Omega})
        \end{bmatrix}\right\}.
\end{align}
The training and testing process follows the same algorithm flow defined in Algorithm \ref{algorithm0}. The RC input dimension is thereby set as $N_r \times N$ (The number of antennas times the axis-length along the Doppler domain). 

\begin{table}[]
\centering
\caption{Notation Appearing in the learning system}
\begin{tabular}{|l|l|l|}
\hline
Notation & Definitions \\ \hline
${\boldsymbol X}_{train} \in {\mathbb C}^{MN_t\times N}$ & Transmitted pilot symbols \\ \hline
${\boldsymbol X}_{test} \in {\mathbb C}^{MN_t\times N}$ & Transmitted data symbols \\ \hline
${\boldsymbol X} \in {\mathbb C}^{MN_t\times N}$ & Transmitted symbols which are assembled in ${\boldsymbol X}_{train}$ and ${\boldsymbol X}_{test} $ \\ \hline 
${\boldsymbol {\mathcal X}} \in {\mathbb C}^{M\times N \times N_t}$& Transmitted symbols in a 3-mode form \\ \hline
${\boldsymbol Y}_{train}  \in {\mathbb C}^{MN_r\times N}$ & Received pilot symbols \\ \hline
${\boldsymbol Y}_{test}  \in {\mathbb C}^{MN_r\times N}$ & Received data symbols \\ \hline
${\boldsymbol Y}  \in {\mathbb C}^{MN_r\times N}$ & Received symbols which contain ${\boldsymbol Y}_{train}$ and ${\boldsymbol Y}_{test}$ \\ \hline
${\boldsymbol {\mathcal Y}} \in {\mathbb C}^{M\times N \times N_r}$& Received symbols in a 3-mode form \\ 
\hline
\end{tabular}
\label{learning_notations}
\end{table}

\subsection{Learning using Multiple RCs}
The aforementioned training framework utilizes one RC in the entire delay-Doppler domain. We can also consider applying multiple RCs in one OTFS frame, where an individual RC is assigned to learn a local channel feature. For instance, when a blockwise pilot pattern is employed for SISO-OTFS, we can split the training dataset (\ref{interleaved_training_set}) into $N$ subsets, where the $n$th subset is defined as
\begin{align}
    \{{\boldsymbol x}_{train}^{(n)} := {\boldsymbol \Omega}_{:,n}\odot {\text {FFT}}({\boldsymbol X})_{:,n},  {\boldsymbol y}_{train}^{(n)} := {\text {FFT}}({\boldsymbol Y})_{:,n}\},
\end{align}
where $\text{FFT}$ operates along the row of the matrix $\boldsymbol X$. Accordingly, the $k$th testing dataset is,
\begin{align}
        \{{\boldsymbol x}_{test}^{(n)} := {\bar{\boldsymbol \Omega}}_{:,n}\odot {\text {FFT}}({\boldsymbol X})_{:,n},  {\boldsymbol y}_{test}^{(n)} := {\text {FFT}}({\boldsymbol Y})_{:,n}\}.
\end{align}
Therefore, the input dimension of each RC is reduced to $1$. We then have $N$ RCs to be individually learned over one OTFS frame. In general, we can group any number of consecutive columns of $\boldsymbol X$ as a subset. Accordingly, the number of RCs is reduced to $K$, where $K$ stands for the number of subsets. Similar to the analysis in \cite{zhou2020learning}, we can easily show that the complexity of using $K$ RCs is proportional to ${\mathcal O}(N_nN_{t}NM+min\{(\eta MN/K)^2, N_n^2\}KN_t)$ FLOPS, where $\eta$ is the percentage of pilot overhead and $N_n$ is the size of the extended state. When $K = N$, the second term in the complexity formula becomes $min\{(\eta M)^2, N_n^2\}NN_t$. When $K = 1$, it is $min\{(\eta M)^2, N_n^2\}N_t$. Therefore, the resulting complexity in these two extreme cases indicates that using multiple RCs tends to be more costly than using one RC in the computation, though the training set size is reduced in the former case. In the next section, we observe that changing the number of RCs can impact the bit error rate performance, offering a tradeoff between the equalization complexity and performance. Meanwhile, the complexity of the message passing (MP) algorithm \cite{raviteja2018interference} is ${\mathcal O}(MNN_t(M\eta N N_t))$, where $M\eta N$ represents the number of non-zero taps of the channel impulse kernel in the delay-Doppler domain. This analysis reveals that MP spends more computational resources than our introduced approach.

\section{Numerical Experiments}
\label{evaluation}
\begin{table}[]
\centering
\caption{Simulation Parameters}
\begin{tabular}{|l|l|}
\hline
\textbf{Parameters} & \textbf{Values} \\ \hline
Channel Model & 3GPP 5G NR CDL \\ \hline
Carrier frequency & 4 GHz       \\ \hline
Delay Profile in Channel Model & CDL-C \\ \hline
Delay Spread & 10 ns \\ \hline
Maximum Doppler Shift & 555 Hz\\ \hline
Carrier Spacing & 15 KHz \\ \hline
Sampling Rate & 15.36 MHz \\ \hline
Pilot Overhead for SISO & 4.69\% \\ \hline
$M$ & 1024 \\ \hline
$N$ & 14 \\ 
\hline
\end{tabular}
\label{sim_para}
\end{table}

In our simulation, we implement OTFS using the standalone approach, as shown in the bottom part of Fig. \ref{implementation_OTFS}. Regarding OTFS parameters, we set $M = 4$ to be compatible with the sub-frame length of OFDM as defined in 4G and 5G systems. The number of sub-carriers is set at 1024. The channel model is selected as 3GPP 5G NR clustered delay line (CDL) channel, where the parameter settings are listed in Table \ref{sim_para}. The CP length for OTFS is set at the maximum channel length generated by the channel model. The channel Doppler shift is 555Hz which represents mobility at highway speeds as high as 90 mph. All curves shown below are obtained by running the algorithms over a transmission link having 30 different time-and-frequency dispersive channel realizations. The training overhead for SISO is set at $4.7\%$ by default, and for MIMO at $18.7\%$. The x-axis is the signal to noise ratio in dB defined as the received signal power normalized by the background noise. 
The RC configuration is based on the paper \cite{zhou2020learning}, where a windowed buffer is added in the input layer with size 20. The size of the recurrent internal state in RC for the superimposed pilot is set as $8$. For the interleaved pilot learned by multiple RCs, the internal state for each RC is set as $8$ for SISO and $12$ for MIMO. Finally, the link reliability metric utilized below is uncoded bit error rate.
 
\begin{figure}
    \centering
    \includegraphics[width = 0.85\linewidth]{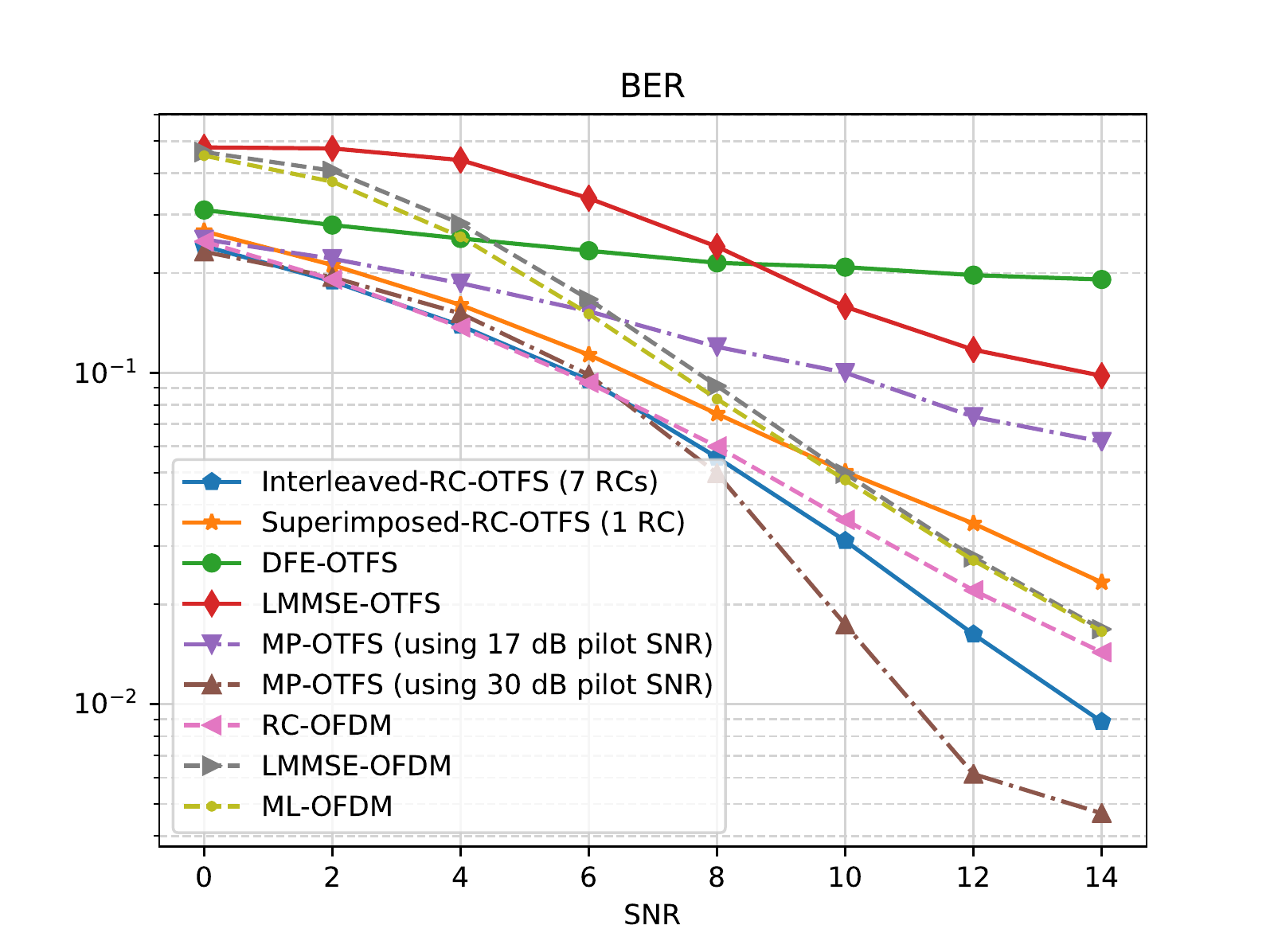}
    \caption{Comparison of BER performance in SISO-OTFS and SISO-OFDM systems with different equalization algorithms. Note that the pilot SNR in MP-OTFS is increased to 17 dB and 30 dB to show its BER drop.}
    \label{OTFS_OFDM_BER}
\end{figure}

The comparison between OFDM and OTFS systems is shown in Fig. \ref{OTFS_OFDM_BER}. The evaluation methods by order of the legend are respectively:
\begin{enumerate}
    \item RC using Interleaved pilot in OTFS (denoted as Interleaved-RC-OTFS), where the number of RCs in one OTFS frame is configured as $7$, and it is selected as the blockwise pattern in Fig. \ref{pilot_design2}.
    \item Superimposed pilot enabled RC using OTFS (denoted as Superimposed-RC-OTFS);
    \item Decision feedback equalization in OTFS using the same amount of training overhead as RC-OTFS (denoted as DFE-OTFS);
    \item LMMSE channel estimation with linear interpolation and LMMSE equalization in time-frequency domain (denoted as LMMSE-OTFS);
    \item Message-passing based equalization in OTFS using estimated CSI which is acquired using the same pilot overhead as other methods. However, the obtained CSI is imperfect in the low SNR regime, which degrades the BER. We show the performance of this approach by increasing the received SNR for pilot symbols to 17 dB;
    \item Message-passing based equalization in OTFS using estimated CSI, where the CSI is obtained by using pilot symbols with SNR 30 dB;
    \item Multiple-RC in OFDM, where 7 RCs are applied in one OFDM subframe. The pilot overhead is set the same as RC-OTFS (denoted as RC-OFDM);
    \item LMMSE channel estimation with linear interpolation and  LMMSE equalization in the time-frequency domain using the same pilot overhead as RC-OTFS (denoted as LMMSE-OFDM);
    \item Maximum likelihood estimation in OFDM with LMMSE channel estimation and interpolation (denoted as ML-OFDM).
\end{enumerate}

\begin{figure}
    \centering
    \includegraphics[width = 0.85\linewidth]{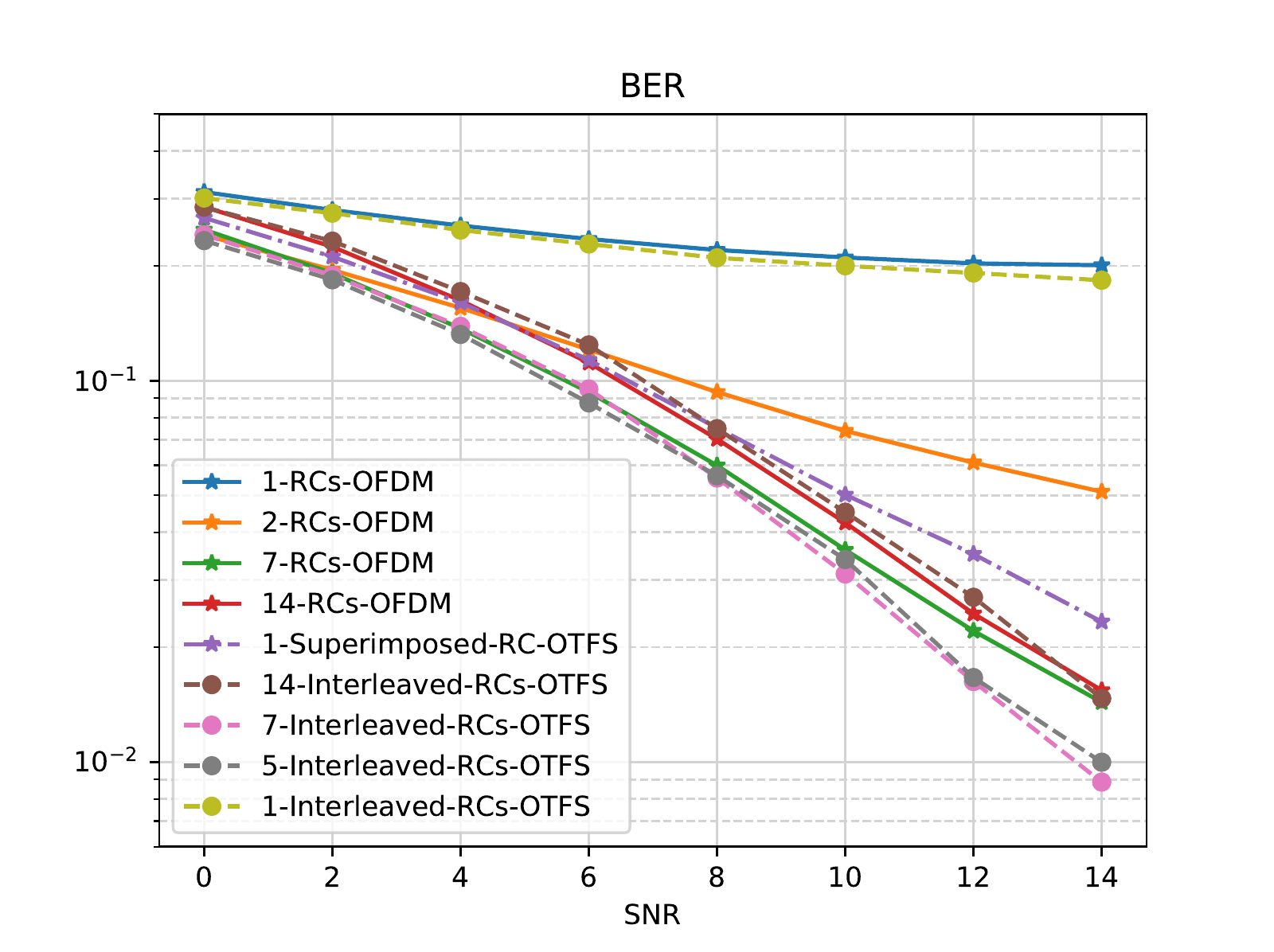}
    \caption{Comparison of BER performance in SISO-OTFS and SISO-OFDM systems with different numbers of RC.}
    \label{RC_SISO_num}
\end{figure}

We observe that RC with OTFS outperforms other approaches in the low SNR regime except for RC-OFDM. The advantage enjoyed by RC-OFDM is because each RC handles the local channel change within each two OFDM symbols. 
To obtain more insight on how the number of RCs utilized in the system can impact performance, we plot the result of using different numbers of RCs in Fig. \ref{RC_SISO_num}. The figure reveals that there is a tradeoff between the number of RCs and the BER performance. In addition, the configuration of each RC in OFDM is the same as the one in OTFS. Moreover, using one RC for the superimposed pilot can significantly improve upon using one for the interleaved pilot. Applying more RCs improves the performance at the cost of increased complexity. As a result, using the OTFS waveform shows incremental gain over OFDM in the evaluation. More significantly, the OFDM system cannot accommodate a superimposed pilot similar to OTFS and achieve a comparable BER performance to OTFS using a single RC. These observations demonstrate the benefits of modulating information symbols in the delay-Doppler domain. 

\begin{figure}
    \centering
    \includegraphics[width = 0.85\linewidth]{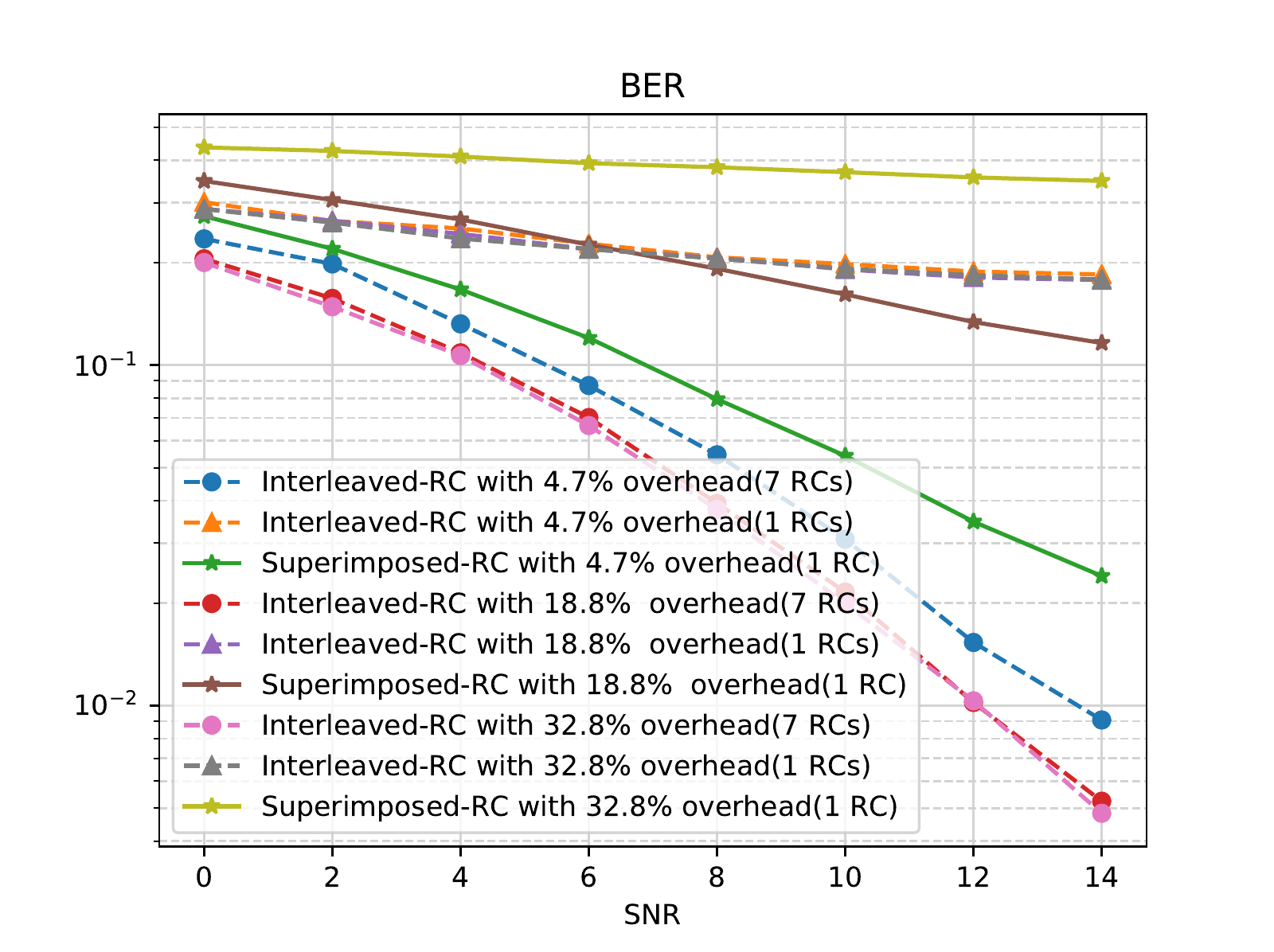}
    \caption{Comparison of pilot overhead for the two different pilot patterns, where the overhead of interleaved RC is calculated in the delay-Doppler domain, while the overhead in Superimposed-RC is calculated in the time-frequency domain.
}
    \label{RC_pilot}
\end{figure}

Fig. \ref{RC_pilot} shows a BER comparison between superimposed pilots and interleaved pilots by changing the overhead. Note that the pilot overhead for the superimposed one is calculated in the time-frequency domain. However, the superimposed pilot-based approach also can be treated as zero overhead in the delay-Doppler domain given a BER threshold. The value of the pilot overhead ranges from $4.7\%$ to $39.0\%$. The scenario under interleaved pilot utilizes 2 RCs in the learning. We can observe that the superimposed pilot becomes inefficient as ``overhead'' increases due to more interference in data symbols added. It is also important to note that the learning approach for interleaved pilots can eliminate the interference between pilot and data symbols even though no guard interval is added. 

\begin{figure}
    \centering
    \includegraphics[height = 0.7\linewidth, width = 0.85\linewidth]{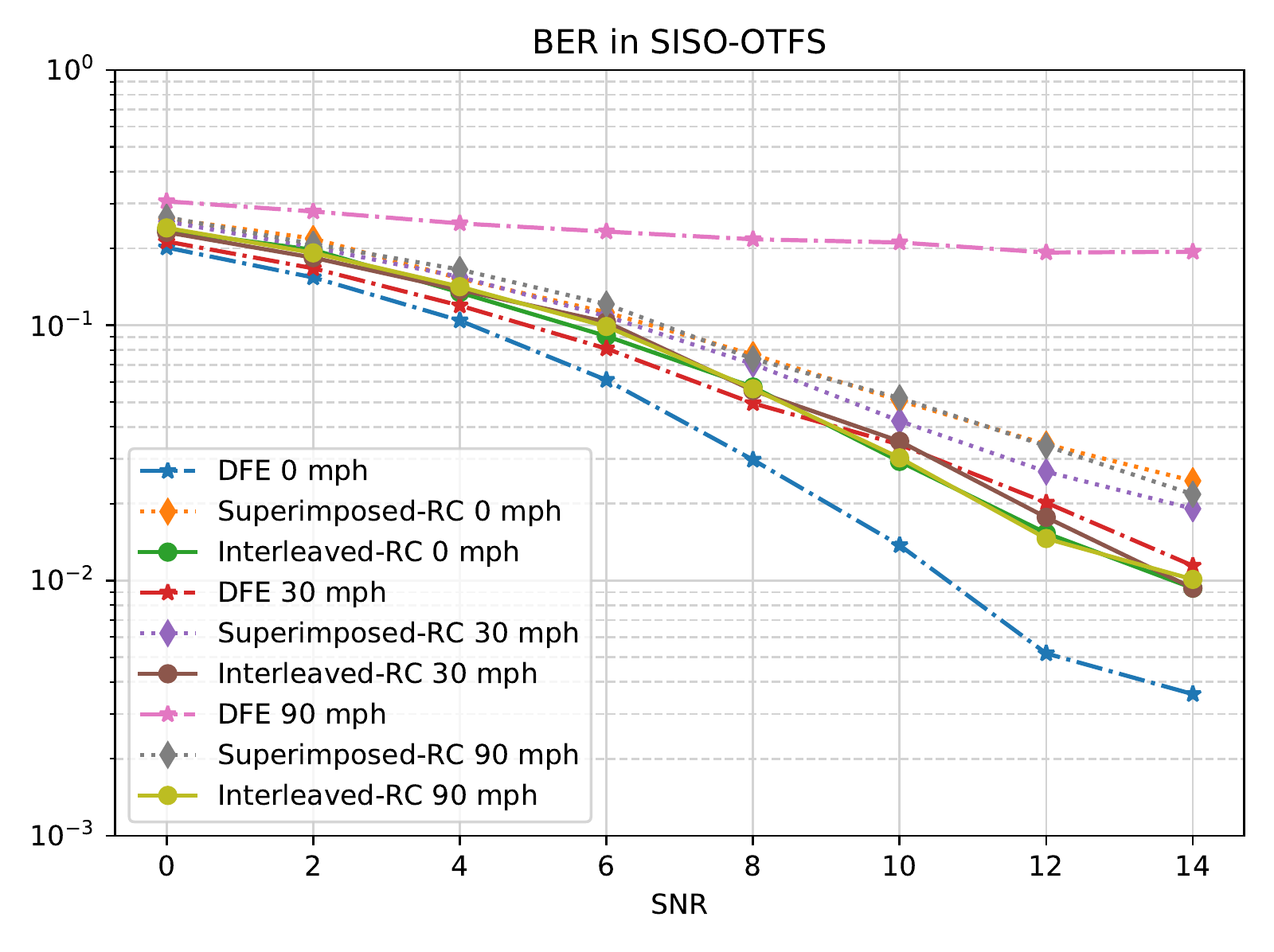}
    \caption{BER in SISO-OTFS under different mobile station (MS) velocities using DFE and RC.}
    \label{rc_doppler}
\end{figure}

\begin{figure}
    \centering
    \includegraphics[height = 0.7\linewidth, width = 0.85\linewidth]{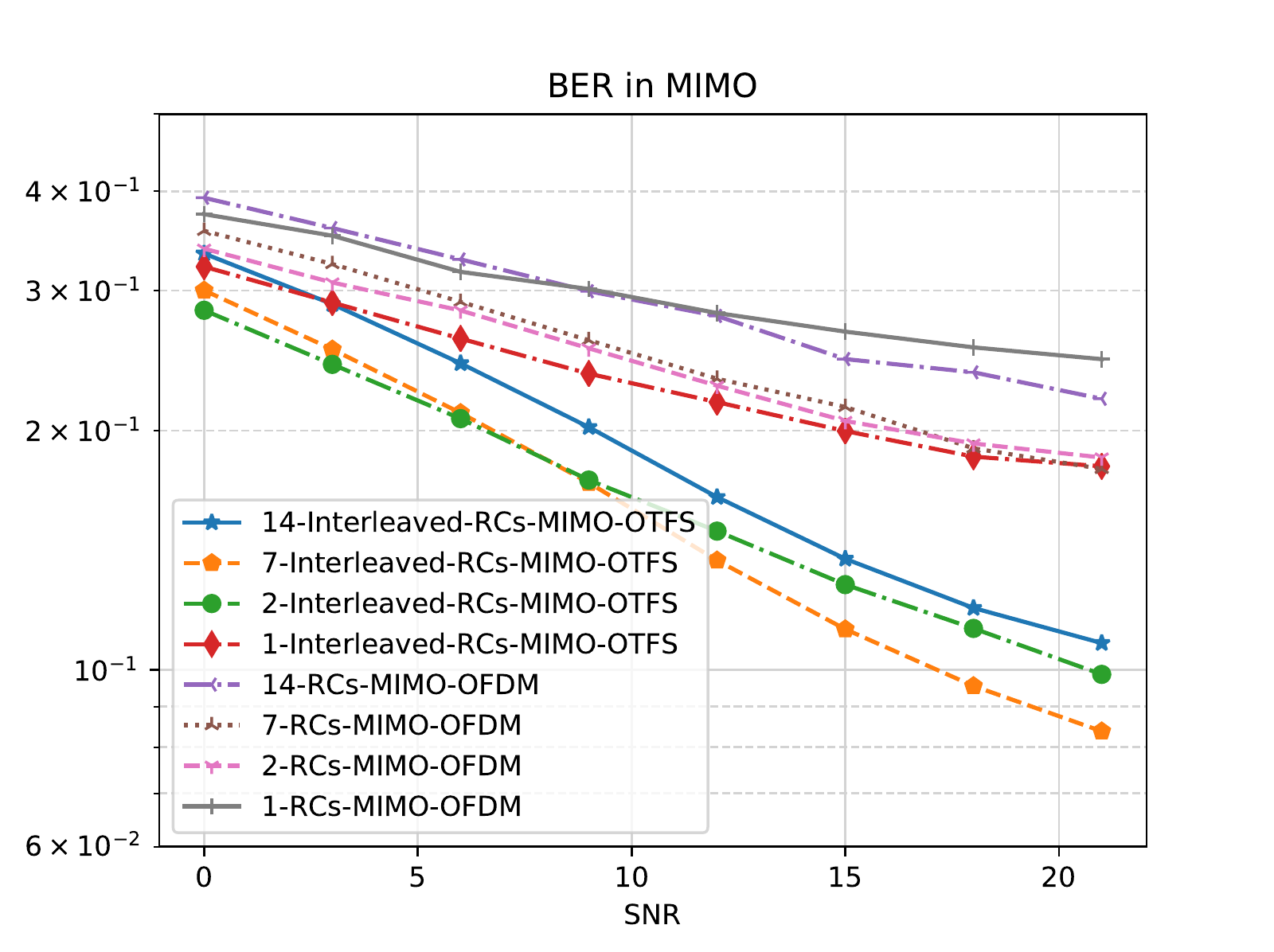}
    \caption{BER of RC in MIMO-OTFS and MIMO-OFDM with different numbers of RC.}
    \label{MIMO_OTFS}
\end{figure}

\begin{figure}
    \centering
    \includegraphics[height = 0.7\linewidth, width = 0.85\linewidth]{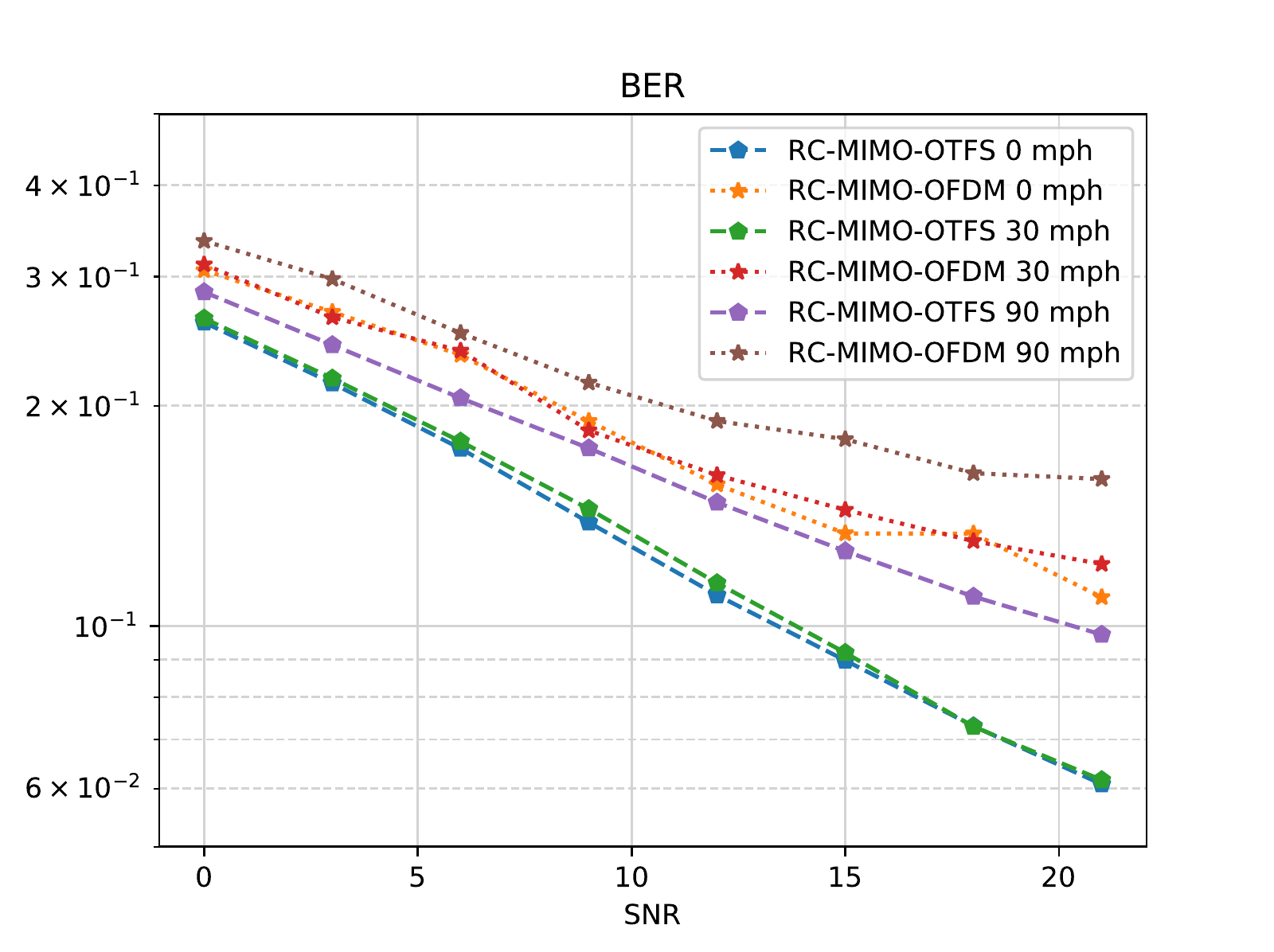}
    \caption{BER of RC in MIMO-OTFS and MIMO-OFDM with different MS velocities.}
    \label{MIMO_OTFS_doppler}
\end{figure}

Fig. \ref{rc_doppler} depicts the BER performance of RC and DFE as we change the value of velocity (Doppler shift). We can observe that the processing of RC is more robust than DFE in the low SNR regime. The poor performance of DFE as SNR increases is because of the imposed limited training constraints (only $4\%$ pilot symbols are utilized via a block pilot pattern in the delay-Doppler domain). To the best of our knowledge, it is still an open problem to efficiently implement DFE in the 2D delay-Doppler domain with a limited length of training symbols. 
Fig.\ref{MIMO_OTFS} shows the BER performance using RC for MIMO-OTFS through blockwise interleaved pilot patterns. We observe that utilizing more RCs need not lead to a monotonic performance improvement. Fig. \ref{MIMO_OTFS_doppler} shows the BER performance under different MS velocities in OTFS and OFDM systems using the same method. The MIMO case also demonstrates the advantage of our approach in the OTFS system. Note that the comparison between the RC-based approach and the conventional OFDM system is studied in \cite{zhou2020learning}. Meanwhile, to the best of our knowledge, model-based approaches in MIMO-OTFS systems have to tackle more interference than the SISO case in the low SRN regime, hence perform far behind the RC learning-based approaches.

\section{Conclusion}
\label{conclusion}
In this paper, we introduced a neural network-based framework for OTFS equalization. Our results demonstrate that using a reservoir computing-based neural network structure coupled with specifically designed pilot patterns (training dataset) can significantly reduce inter-symbol interference. More significantly, the training of the equalization neural network utilizes observations from a single OTFS frame which is competitive with the control plane employed in current cellular systems. The success of the resulting ``learning through one OTFS frame'' is also because the offered channel is an invariant feature of OTFS modulation, which transforms doubly selective wireless channels to nearly constant convolution kernels in the delay-Doppler domain. Our numerical results show that the neural network-based equalization method outperforms traditional methods in various Doppler spread settings. In addition, we can observe the performance gain of OTFS by applying the same neural network structure to an OFDM system with the same pilot control overhead. In conclusion, we believe an OTFS system employed with our implemented equalizer can provide either better transmission reliability or lower processing complexity to state-of-the-art equalization approaches for OFDM systems in high Doppler environments. 

\bibliographystyle{IEEEtran}
\bibliography{IEEEabrv,./ref}


\renewenvironment{IEEEbiography}[1]
  {\IEEEbiographynophoto{#1}}
  {\endIEEEbiographynophoto}

\end{document}